\documentclass[journal]{IEEEtran} 
\def\mycmd{0} % 0: 2 cols/ else: 1 column. Setear la linea de arriba en concordancia con esto.
\usepackage{cite}
\usepackage[ruled,vlined]{algorithm2e}
\usepackage[english]{babel}
\usepackage{epsfig}
\usepackage{color}
\usepackage{times}
\usepackage{subfigure}
\usepackage{url}
\usepackage{amsthm}
\usepackage{amsmath}
\usepackage{amssymb}
\usepackage{dsfont}
\usepackage{stackengine}
\usepackage{acronym}
\theoremstyle{definition} % \normalfont en los teoremas, etc
\newtheorem{lemma}{Lemma} %[section]
\newtheorem{theorem}{Theorem}
\newtheorem{remark}{Remark} %[section]
\newcommand{\Hip}{\mathcal{H}}

\newcommand{\CN}{\mathcal{CN}}

\newcounter{mycomment}

\begin{document}
	\title{An Exponentially-Tight Approximate Factorization of the Joint PDF of Statistical Dependent Measurements in Wireless Sensor Networks}
	
	\author{{Juan Augusto Maya, Leonardo Rey Vega and Andrea M. Tonello,~\IEEEmembership{Senior Member,~IEEE.}}
		\thanks{J. A. Maya and A. M. Tonello are with the University of Klagenfurt, Klagenfurt, Austria. L. Rey Vega is with Universidad de Buenos Aires, Buenos Aires, Argentina, and also with CSC-CONICET, Buenos Aires, Argentina.
			The work of J. A. Maya has been supported by the Ubiquitous Sensing Lab, a joint laboratory between the University of Klagenfurt and Silicon Austria Labs. The work of L. Rey Vega was supported by grant PICT 2020-Serie A-01336.
			Emails:  juan.maya@fi.uba.ar, lrey@fi.uba.ar, andrea.tonello@aau.at.
			%This is a pre-print version of an article submitted to IEEE Transactions on Wireless Communications and it is under review.
		} %
	}
	
	\maketitle
	
	\begin{abstract}
		We consider the distributed detection problem of a temporally correlated random radio source signal using a wireless sensor network capable of measuring the energy of the received signals. It is well-known that optimal tests in the Neyman-Pearson setting are based on likelihood ratio tests (LRT), which, in this set-up, evaluate the quotient between the probability density functions (PDF) of the measurements when the source signal is present and absent. When the source is present, the computation of the joint PDF of the energy measurements at the nodes is a challenging problem. This is due to the statistical dependence introduced to the received signals by the propagation through fading channels of the radio signal emitted by the source. We deal with this problem using the characteristic function of the (intractable) joint PDF, and proposing an approximation to it. We derive bounds for the approximation error in two wireless propagation scenarios, slow and fast fading, and show that the proposed approximation is exponentially tight with the number of nodes when the time-bandwidth product is sufficiently high. The approximation is used as a substitute of the exact joint PDF for building an approximate LRT, which performs better than other well-known detectors, as verified by Monte Carlo simulations.
	\end{abstract}
	
	%IEEE Open Journal of the Communications Society
	\section{Introduction}
	Wireless sensor networks (WSNs) as a key technology in the emerging paradigm of Internet of Things (IoT) \cite{gubbi2013internet,al2015internet,gupta2020collaborative} have received considerable attention. Distributed signal processing is an important topic of research in this area, because efficient information processing in large networks of devices with limited communication, sensing, storage and computing capabilities has the potential of being cost-efficient and very robust \cite{DiLorenzo_Barbarossa_Sardellitti_2020}. Among the different signal processing tasks in WSNs, distributed detection is one of the most important \cite{chepuri2016sparse,ciuonzo2017distributed, aldalahmeh2019fusion}.
	
	In the distributed detection problem, a set of nodes sense the environment in search for the presence of a source signal, which is typically linked with some physical process extended over the geographical area where the network is deployed. Through collaboration among the nodes, the network is expected to decide with high confidence if the above mentioned signal is present or not. A well-studied application of this general problem is the spectrum sensing task in cognitive radios \cite{Lunden_Koivunen_Poor_2015}. An important issue is the presence and influence of the spatial and temporal correlation of the source signal on the implemented detection scheme. The source signal is sometimes modeled as a stochastic process which could present temporal correlation (e.g. cyclostationarities, \cite{Lunden_Koivunen_Huttunen_Poor_2009}). On the other hand, when the source signal is present, the measurements taken at each sensor node are clearly correlated (and therefore, statistically dependent) because these measurements are different noisy versions (affected by channel effects such as path-loss, shadowing, and fading) of the same random signal.
	
	\subsection{Motivation and some previous works}
	
\begin{figure}
		\centering
		\if\mycmd0
		\includegraphics[width=\linewidth]{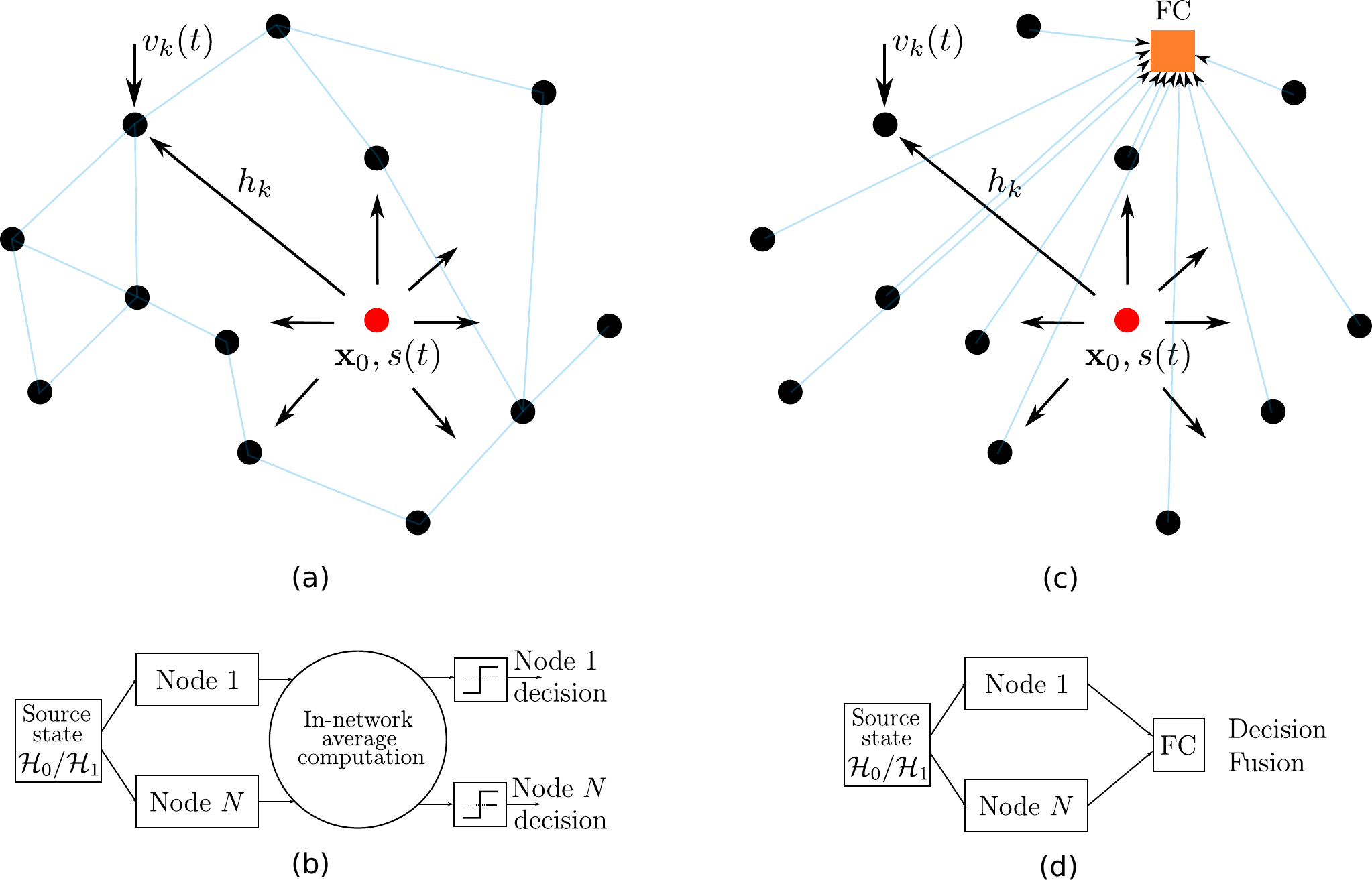}
		\else
		\includegraphics[width=.5\linewidth]{network.pdf}
		\fi
		\caption{Two common network architectures. (a) Fully-distributed scenario. All nodes (black dots) compute a local statistic, exchange information through short-range communication links (light-blue lines), and lastly, build the final test statistic to make a decision. (b) A block diagram for the scenario in (a). (c) The network has a FC (orange square), where all the (quantized) measurements are sent from the nodes and the final decision is made. (d) A block diagram for the scenario in (c).}
		\label{fig:network}
	\end{figure}

	It is well-known that the optimal test \cite{Levy_Det}, which can take into account the correlation (or more generally, the statistical dependence) of the signals, is the likelihood ratio test (LRT), which is defined through the quotient between the probability density functions (PDF) of the measurements when the signal is present and not. However, in general, the exact implementation of the optimal LRTs is difficult in fully-distributed settings, where there is no fusion center (FC), and the cooperation between different sensor nodes is done through transmissions between neighbor nodes. See Fig. \ref{fig:network} (a) and (b) for a schematic representation of the problem. The reason is that those tests require network-wide interactions between the measurements taken at the sensor nodes. For example, when the signals captured at each sensor site are Gaussian, the optimal tests involve the computation of quadratic forms of the measurements. In fully-distributed consensus detection schemes \cite{Maya_2021,al2018node,sayed2013diffusion}, this is problematic when the covariance matrix of the measurements is not diagonal. The number of exchanges between the nodes in the network needed to construct the consensus decision could be large, which introduces severe penalties in power and bandwidth consumption, which are typically scarce resources in WSNs composed of typically inexpensive nodes. When a FC is present (see Fig. 1 (c) and (d)), the sensors could transmit their measurements to the FC, which is responsible of computing the final decision about the presence of the source signal \cite{willett2000good, li2018fully, Ciuonzo_Rossi_Willett_2017}. This centralized architecture also presents some
	weaknesses as, for example, it lacks of robustness against
	the malfunctioning of a single device, given that a failure in
	the FC may severely degrade the performance of the system. Additionally, it requires that sensor nodes, typically battery powered devices, communicate through orthogonal channels with the FC, in which case the energy and bandwidth resources
	increase with the amount of sensors and the area in which the network is deployed. This could be alleviated, in principle, using  fusion decision algorithms \cite{ciuonzo2014decision,Li_Li_Varshney_2020,Ciuonzo_Rossi_Varshney_2021,Ahmadi_Maleki_Vosoughi_2018,rossi2016performance}, where the local measurements (or a statistic of them) are quantized to a few bits. These decisions are then communicated to a FC, where a typically suboptimal final test statistic is built and the final decision is made. However, the influence of spatial/temporal correlation in the construction of the local and final test statistic is not usually taken into account for their design, not modeled, or even discarded without an analytical justification.
	
	\subsection{Our contributions and paper organization}
	
	Instead of restricting us to a specific architecture (with or without a FC), we begin in Section \ref{sec:problem} formulating the problem of detecting an arbitrary temporal-correlated random source signal. We consider that the source signal and the additive noise at each sensor are Gaussian distributed signals, and that the sensors deliver energy measurements.
	Because the source signal is modeled as a random signal, the node’s observations are statistically dependent, given that each sensor receives a noisy version of the same source signal propagated through the corresponding wireless channel.
	The exact PDF of the energy measurements at the sensor sites when the source signal is present, which corresponds to the diagonal entries of a Wishart-distributed matrix, is not known in closed form \cite{Royen_1991}. Although there exist some recent results \cite{Morales-Jimenez_Paris_Entrambasaguas_Wong_2011}, \cite{Ermolova_Tirkkonen_2012} in terms of multidimensional series that involve Laguerre polynomials, the expressions are not particularly easy to handle and compute (specifically in high-dimensional scenarios, i.e., many nodes, due to numerical precision issues), and they do not cover the more general case considered in this paper (the arbitrarily temporally correlated source signal case). From this starting point, our main contribution is the derivation of a tight approximation to the actual joint PDF of the energy measurements, which conveniently factorizes itself in the product of $N$ (the number of sensors) PDFs, each of one is parameterized through parameters that each node can compute or obtain locally. This approximation is treated in two typical and important propagation scenarios: slow and fast fading. This finding presents two major advantages in distributed detection scenarios: i) It allows us to build simpler cooperative schemes for detecting a radio source by using any of the two network architectures shown in Fig. \ref{fig:network}. ii) If desirable, it is possible to design simpler quantization schemes by relying on the local PDF in each node instead of the intractable joint PDF.
	
	In Section \ref{sec:approx}, we calculate in closed form the characteristic function (CF) of the energy measurements. From this exact CF, and under the hypothesis of a high time-bandwidth product $WT$, which is a common situation in several practical applications, the approximation to the true joint PDF of the energy measurements is derived. This product PDF is non-trivial (i.e., it is not the product of the marginals of the true joint PDF) and can be used as a replacement of the true non-product joint PDF for implementing the LRT or quantization schemes, among others. In addition, this result implies that, when $WT$ is large and for energy-based detectors, the temporal and spatial statistical dependence of the measurements are not critical in this set-up.
	
	In Section \ref{sec:analysis}, a theoretical study of the obtained approximation for the true PDF is also provided for both above-mentioned scenarios, showing the quality of the approximation in terms of the value of $WT$ and the network size $N$. In this respect, the most important fact is that the approximation is exponentially tight with $N$. In Section \ref{sec:test_approx}, in order to cope with the unknown remaining parameters in the obtained approximations, we obtain the corresponding generalized likelihood ratio tests (GLRT). Finally, in Section \ref{sec:numerical_results}, some numerical simulations are also conducted to evaluate the performance of the obtained tests for detecting the presence of a radio source with unquantized energy measurements. The paper is finalized with some concluding remarks and future research directions. The proofs of some mathematical results are relegated to the appendices.
	
	This work is an extension of our conference paper \cite{maya2020cf}, where only preliminary results were presented. Specifically, we here provide new results for the fast fading scenario, the theoretical error approximation analysis and new simulations for both fast and slow fading scenarios. We also include the mathematical proofs of the results.   
	
	\subsection{Notation and acronyms}
	Vectors/matrices are denoted by lowercase/uppercase boldface. Also, depending on the context, with uppercase boldface letters we denote random vector and matrices. With $\mbox{diag}(a_1,a_2,\dots, a_N)$ we denote the diagonal $N\times N$ matrix with diagonal entries given by $a_1,a_2,\dots, a_N$. The trace and the determinant of square matrix $\mathbf{A}$ are denoted by $\mbox{Tr}(\mathbf{A})$ and $|\mathbf{A}|$, respectively. For a complex number $z\in\mathbb{C}$, we denote its magnitude with $|z|$. The use of $|\cdot|$ for the determinant of a square matrix or for the magnitude of complex number will be clear from the context. The Kronecker product between matrices $\mathbf{A}$ and $\mathbf{B}$ is denoted as $\mathbf{A}\otimes\mathbf{B}$. A vector $\mathbf{y}$ complex circular Gaussian distributed with mean $\mathbf{a}$ and covariance matrix $\mathbf{B}$ has PDF denoted by $\mathbf{y}\sim\mathcal{C}\mathcal{N}(\mathbf{a},\mathbf{B})$. The $N\times N$ identity matrix is denoted as $\mathbf{I}_N$.  Symbols $P_{\text{fa}}$ and $P_{\text{md}}$ denote  the false-alarm probability and miss-detection probability of the statistical detection test considered. We use the big-O notation, that is, $f(x)=\mathcal{O}(g(x))$ as $x\rightarrow\infty$ and for $g(x)$ a strictly positive function if $|f(x)|\leq K g(x)$ when $x>x_0$ for $K$ and $x_0$ positive constants. The symbol $\equiv$ is used to introduce the definition of a new quantity.
	
	%\subsection{Acronyms}
	For easier reference, we list some common acronyms used in the text:
	\begin{acronym}
		%\small
		\acro{CF}{Characteristic Function}
		\acro{CSI}{Channel State Information}  
		\acro{CROC}{Complementary Receiver Operating Characteristic}  
		\acro{FC}{Fusion Center}
		\acro{GLRT}{Generalized Likelihood Ratio Test}
		\acro{IoT}{Internet of Things}   
		\acro{LRT}{Likelihood Ratio Test}  
		\acro{MLE}{Maximum Likelihood Estimator}  
		\acro{PDF}{Probability Density Function}
		\acro{SNR}{Signal to Noise Ratio}
		\acro{UMPT}{Uniformly Most Powerful Test}  
		\acro{WSN}{Wireless Sensor Network}
	\end{acronym}
	
	\section{Problem setting}
	\label{sec:problem}
	Consider the problem of $N$ sensors distributed in a given geographical area. All of them have sensing and communication capabilities. It is assumed that a radio source is located in the same area. The location of this radio source is unknown for the sensing network, and it is assumed that it emits an arbitrarily temporally-correlated Gaussian signal. Each node of the network senses the environment during $L$ time-windows, each of duration $M$ symbols. During the $l$-th time-window (with $l\in[1:L])$, the received signal at each sensor $n\in[1:N]$ can be written as:
	\begin{equation}
		\label{eq:signal_model_1}
		\mathbf{y}_{n,l}=h_{n,l}\mathbf{s}_l+\mathbf{v}_{n,l},
	\end{equation}
	where for each $n,l$, $\mathbf{y}_{n,l}\in\mathbb{C}^M$ is the temporal baseband signal at sensor $n$ and time-window $l$, which is composed of  the noise signal $\mathbf{v}_{n,l}\in\mathbb{C}^M$, and the source signal  (when present) $\mathbf{s}_{l}\in\mathbb{C}^M$ . This source signal is affected by the channel gain $h_{n,l}\in\mathbb{C}$ which includes, among other effects, the path-loss gain and fading that exists between the position of the source and the $n$-th sensor position at measurement window $l$. This coefficient, is highly dependent on the scenario under study (i.e, a high mobility situation or a stationary one) and it is typically unknown in practice or difficult and too complex to acquire in scenarios where the source does not cooperate to be detected. Notice that, as the signal $\mathbf{s}_{l}$ is the same for each sensor (although affected by different channel coefficients $h_{n,l}$), the signals $\{\mathbf{y}_{n,l}\}_{n=1}^N$ received at each sensor node are spatially correlated for each measurement window $l$. Besides that, as the radio source signal $\mathbf{s}_{l}$ can also be temporally correlated, the signals $\mathbf{y}_{n,l}$ can also present temporal correlation.
	
	It is assumed that the sensors have limited sensing capabilities and they can only sense the average energy during each measurement window. This is a practical and typical situation in several applications such as cognitive radio \cite{Liang_Zeng_Peh_Hoang_2008}, and it was extensively studied  \cite{Unnikrishnan_Veeravalli_2008}, \cite{digham2003energy}.  
	Energy detectors are typically available even in inexpensive hardware and they do not require coherent demodulation procedures, in which prior information related to the radio source's waveform or modulation is needed.
	Besides that, the optimal detection test for stochastic Gaussian signals (\ref{eq:signal_model_1}) has the well-known estimator-correlator structure that leads to a \emph{weighted} energy decision statistic \cite{Kay_SSP}, \cite{Kailath_Poor_1998}.
	Therefore, dealing with the case where the nodes deliver only energy measurements is reasonable, from a practical and a theoretical point of view. So, for each time-window $l\in[1:L]$, each sensor $n\in[1:N]$, outputs the value $E_{n,l}\equiv\frac{1}{\beta(M)}\|\mathbf{y}_{n,l}\|^2$, where $\beta(M)>0$ is an appropriate normalizing constant, that depends on the length of $\mathbf{y}_{n,l}$. This constant is increasing with $M$ and some usual choices could be $\beta(M)=M$ or $\beta(M)=M^{1/2}$. These energy values are then used, through the following hypothesis test, by the sensing network to determine if the radio source is actually transmitting or not:
	\begin{equation}
		\left\{\hspace{-1mm}
		\begin{array}{l}
			\mathcal{H}_0: E_{n,l}=\tfrac{1}{\beta(M)}\|\mathbf{v}_{n,l}\|^2,\ n\in[1:N],\ l\in[1:L]\\
			\mathcal{H}_1: E_{n,l}=\tfrac{1}{\beta(M)}\|h_{n,l}\mathbf{s}_l+\mathbf{v}_{n,l}\|^2.
		\end{array}\right.
		\label{eq:energy_test}
	\end{equation}
	It is well-known that the likelihood ratio between the PDFs under both hypotheses is a key element for optimal detection strategies. We denote with $\mathbf{E}_l=\left[E_{1,l},E_{2,l},\dots, E_{N,l}\right]^T$ the energy measurements at each sensor site during time-window $l$, and with $p_{l}^{0}(\mathbf{E}_l)$ and $p_{l}^{1}(\mathbf{E}_l)$ the PDFs under $\mathcal{H}_0$ and $\mathcal{H}_1$, respectively. We assume that the energy measurements $\{\mathbf{E}_l\}_{l=1}^L$ are independent for different time windows $l$\footnote{In practical scenarios, if $M$ is set large enough, the temporal statistical dependence of the source signal between different time-windows vanishes.}. The optimal decision statistics in the Neyman-Pearson sense is \cite{Levy_Det}:
	\begin{equation}
		\log\frac{\prod_{l=1}^L p_{l}^{1}(\mathbf{E}_l)}{\prod_{l=1}^Lp_{l}^{0}(\mathbf{E}_l)}\underset{\mathcal{H}_0}{ \overset{\mathcal{H}_1}{\gtrless}} \tau,
		\label{eq:LR}
		\end {equation}
		where $\tau$ is a threshold usually chosen to set the false alarm probability $P_{\text{fa}}\equiv\mathbb{P}\left\{\mbox{decide for}\ \mathcal{H}_1|\mathcal{H}_0\right\}$. The miss-detection probability is defined by $P_{\text{md}}\equiv\mathbb{P}\left\{\mbox{decide for}\ \mathcal{H}_0|\mathcal{H}_1\right\}$. For this reason and in order to analyze this hypothesis testing problem, we first need the joint PDF, over the sensing nodes, of the energy measurements $\mathbf{E}_l$, for each measurement window.  
		
		In order to closely investigate the PDFs under both hypotheses we will make some standard assumptions. Under $\mathcal{H}_0$ (the radio source is not emitting), the signal $\mathbf{v}_{n,l}$ is a complex Gaussian circular noise, that is, for each $n,l$, $\mathbf{v}_{n,l}\sim\mathcal{C}\mathcal{N}(\mathbf{0},\sigma_v^2\mathbf{I}_M)$, $\sigma_v^2>0$. When the  radio source is emitting (under $\mathcal{H}_1$), the signal at each sensor comprises the noise signal $\mathbf{v}_{n,l}$ with the same characteristics as before, and also the signal $h_{n,l}\mathbf{s}_l$. It is assumed that $\mathbf{s}_l$ is a complex and circular Gaussian signal distributed as $\mathbf{s}_{l}\sim\mathcal{C}\mathcal{N}(\mathbf{0},{\bf\Sigma}_{\mathbf{s},l})$, independent of the noise signal, and where ${\bf\Sigma}_{\mathbf{s},l}\in\mathbb{C}^{M\times M}$ is the temporal correlation matrix of the radio source signal during the measurement window $l$.
		
		Although the Gaussian assumption may seem restrictive, it is widely used, even in situations where it is not strictly true. This is the case of, for example, cognitive radio applications, where the radio source's waveform is an orthogonal frequency-division multiplexing (OFDM) signal, with symbols taking values from a discrete set (e.g., QAM). In this example, the signal presents temporal correlation due to the cyclic prefix introduced by OFDM. We include numerical experiments of this scenario in Section VI. Moreover, as it is shown in the next section, the Gaussian assumption will permit us to obtain closed-form expressions of several important quantities and to deal with the spatial and temporal correlation of the measurements in the above detection problem.
		
		{The model assumed above and the results in Sections III and IV are, in principle, agnostic of the considered fusion setup (fully-distributed or centralized setup with a FC). Depending on the chosen fusion setup, the main differences will be in the demanded network communications resources to compute the test statistics derived from our PDF approximation.}
		\begin{remark}
			It is straightforward to show that the performance of the LRT is invariant to the scaling coefficient $\beta(M)$ in (\ref{eq:energy_test}).  
			That is, for a given value of $P_{\text{fa}}$, $P_{\text{md}}$ does not depend on $\beta(M)$. So, the assumed scaling coefficient has no practical consequence to the implementation of the LRT in (\ref{eq:LR}). However, the scaling coefficient will be pivotal in Section \ref{sec:approx} and \ref{sec:analysis}, where we will analyze the influence of the spatial dependence in $\mathbf{E}_l$ under $\mathcal{H}_1$, when $M$ is large.
		\end{remark}
		
		\section{On the approximation of the PDF of $\mathbf{E}_l$}
		\label{sec:approx}
		In this section, we first study the PDFs of the energy measurements under both hypotheses.  We consider two scenarios:
		\paragraph{The slow fading scenario} The channel gains between the source signal and sensor sites are fixed for all  measurement windows, i.e, $h_{n,l}\equiv h_n$  $\forall l\in[1:L]$ and $n\in[1:N]$. This is representative of a stationary environment.
		\paragraph{The fast fading scenario} The channel gains between the source signal and sensor sites change for each measurement window according to a given distribution (e.g, Rayleigh distribution). This is representative of a high-mobility environment.
		
		For these two scenarios, as obtaining the joint PDF of the measurements is a challenging problem, the study of the CF of $\mathbf{E}_l$ conditioned on the values of  channels gains between the source signal and the sensors proves to be critical. For that reason, we will proceed by obtaining a closed form expression of this CF and use it to obtain and mathematically justify our approximations.  In the following, and except when this will be strictly needed to avoid confusion, we will consider a fixed measurement window with $l\in[1:L]$ of size $M$ and we will omit the subscript $l$ in all relevant quantities in order to use a less cumbersome notation. We will also define the vector of channel gains $\mathbf{h}=\left[h_{1},h_{2},\dots,h_{N}\right]^T$.  We will begin analyzing the CF of $\mathbf{E}$ for the slow fading scenario which is mathematically equivalent to the CF of $\mathbf{E}$ conditioned on $\mathbf{h}$.
		
		\subsection{On the characteristic function of $\mathbf{E}$ conditioned on $\mathbf{h}$}
		
		We define the vector $\mathbf{y}\equiv\left[\mathbf{y}_{1}^T,\mathbf{y}_{2}^T,\dots, \mathbf{y}_{N}^T\right]^T$ as the $NM$-length vector that contains the signals during the considered measurement window at the $N$ sensor positions. Clearly, $\mathbf{y}\sim\mathcal{C}\mathcal{N}(\mathbf{0},\sigma_v^2\mathbf{I}_{NM})$ under $\mathcal{H}_0$. It is also not difficult to show that $\mathbf{y}\sim\mathcal{C}\mathcal{N}(\mathbf{0},\mathbf{h}\mathbf{h}^H\otimes{\bf \Sigma}_{\mathbf{s}}+\sigma_v^2\mathbf{I}_{NM})$ under $\mathcal{H}_1$. The exact joint density of the vector $\mathbf{E}$ is a very difficult problem \cite{Jensen_1970}, \cite{Royen_1991}. It is well-known that it is related to the distribution of the diagonal of $\mathbf{Y}\mathbf{Y}^H$, where $\mathbf{Y}$ is the vertical concatenation of vectors $\mathbf{y}_{n}^H$, $n\in[1:N]$. Although this problem has attracted some interest in the wireless communications community (see \cite{Morales-Jimenez_Paris_Entrambasaguas_Wong_2011} and \cite{Ermolova_Tirkkonen_2012}), most of the results consider the case in which the vectors $\mathbf{y}_{n}$, $n\in[1:N]$ are independent and identically distributed (i.i.d.), which is not the case considered in this paper.
		The main reason behind this is the presence of the channel coefficient between source and sensor locations. This naturally generates a statistical dependence between vectors $\mathbf{y}_{n}$, $n\in[1:N]$ which is difficult to characterize and quantify. If in addition, the radio source signal presents temporal correlation, this will also contribute to the mentioned dependence. Most solutions, in the i.i.d. case, consider inverse Fourier methods and series expansions for the joint density. This is motivated by the fact that the  CF for the diagonal of  $\mathbf{Y}\mathbf{Y}^H$ can be easily computed in that case.
		
		In this paper, we will perform the exact computation of the CF of $\mathbf{E}$, taking into account the specific signal model defined in the previous section which does not match the usual characteristics of the i.i.d. signal model assumed in most of the literature. The obtained CF will then be used, not for an exact series expansion for the PDF of $\mathbf{E}$, but to obtain a meaningful approximation to the PDF. As we will see, this approximation has very useful properties for the distributed detection problem in (\ref{eq:energy_test}). Consider the CF of $\mathbf{E}$ conditioned to $\mathbf{h}$ and the hypothesis $\mathcal{H}_i$: $\Psi^{i}({\boldsymbol\omega}|\mathbf{h})\equiv \mathbb{E}\left[e^{j{\boldsymbol\omega}^T\mathbf{E}}\Big|\mathbf{h},\Hip_i\right]$
		where $\boldsymbol{\omega}\in\mathbb{R}^N$ and $i=0,1$ denotes the true state of nature (i.e, $\mathcal{H}_0$ or $\mathcal{H}_1$). The following lemma, proved in Appendix \ref{ap:CF}, gives us the exact result for $\Psi^{i}({\boldsymbol\omega}|\mathbf{h})$.
		
		\begin{lemma}
			The CF of $\mathbf{E}$ conditioned to $\mathbf{h}$ and the hypotheses $\mathcal{H}_0$ and $\mathcal{H}_1$ are\footnote{Strictly speaking the characteristic function under $\mathcal{H}_0$ does not depend on $\mathbf{h}$. However, in order to use a uniform notation across the paper we will continue using $\Psi^0(\boldsymbol{\omega}|\mathbf{h})$.}, respectively,
			\if\mycmd1
			`
			\else
			\begin{align}
				\Psi^0(\boldsymbol{\omega}|\mathbf{h})=&\prod_{n=1}^N \left(1-j\frac{\omega_n\sigma_v^2}{\beta(M)}\right)^{-M}, \label{eq:CF_cero}\\
				\Psi^1(\boldsymbol{\omega}|\mathbf{h})=& \frac{\prod_{n=1}^N\left(1-j\frac{\omega_n\sigma_v^2}{\beta(M)}\right)^{-M}}{\prod_{m=1}^M\left(1-j\frac{\lambda_{m}}{\beta(M)}\sum_{n=1}^N\frac{|h_{n}|^2\omega_n}{1-j\frac{\omega_n\sigma_v^2}{\beta(M)}}\right)},\label{eq:CF_uno}
			\end{align}
			\fi
			where $\lambda_{1},\lambda_{2},\dots,\lambda_{N}$ are the eigenvalues of the source signal covariance matrix $\boldsymbol{\Sigma}_{\mathbf{s}}$.
			\label{lemma:CF}
		\end{lemma}
		
		The previous lemma deserves some comments. In the first place, under $\mathcal{H}_0$, the energy vector density is distributed as $N$ identical and independent central chi-square random variables with $2M$ degrees of freedom and mean $\frac{M\sigma_v^2}{\beta(M)}$.  Moreover, as the source is not transmitting, neither the CF or PDF depend on $\mathbf{h}$. Under $\mathcal{H}_1$, $\Psi^1(\boldsymbol{\omega}|\mathbf{h})$ can not be factored as $\prod_{n=1}^N\psi_{n}(\omega_n|\mathbf{h})$, where each $\psi_{n}$ with $n\in[1:N]$ is a characteristic function. As expected, $\{E_{n}\}$ are dependent random variables independently of the time correlation characteristics of the source signal $\mathbf{s}$. From (\ref{eq:CF_uno}), this is true even when the source signal is uncorrelated in time, i.e. $\lambda_m=\sigma_s^2$ with $m\in[1:M]$. This is a consequence of the spatial correlation induced by the common source signal present at each sensor site.
		
		In principle, there is no known closed form expression for the joint density of $\mathbf{E}$. However it is easy to show, assuming $\lambda_{i}\neq\lambda_{j,}$ with $i\neq j$, that the marginal density for $E_{n}$ with $n\in[1:N]$ can be written as:
		\if\mycmd1
		\begin{gather}
			p^1(E_{n}|\mathbf{h})=\sum_{m=1}^M\prod_{j\neq m}^M\left(1-\frac{\lambda_{j}|h_{n}|^2+\sigma_v^2}{\lambda_{m}|h_{n}|^2+\sigma_v^2}\right)^{-1}\frac{\beta(M)}{\lambda_{m}|h_{n}|^2+\sigma_v^2}\exp\left(-\frac{\beta(M)E_{n}}{\lambda_{m}|h_{n}|^2+\sigma_v^2}\right),\ \ n\in[1:N].
			\label{eq:marginal_E}
		\end{gather}
		\else
		\begin{gather}
			p^1(E_{n}|\mathbf{h})=\sum_{m=1}^M\prod_{j\neq m}^M\left(1-\frac{\lambda_{j}|h_{n}|^2+\sigma_v^2}{\lambda_{m}|h_{n}|^2+\sigma_v^2}\right)^{-1}\frac{\beta(M)}{\lambda_{m}|h_{n}|^2+\sigma_v^2}\nonumber\\
			\times \exp\left(-\frac{\beta(M)E_{n}}{\lambda_{m}|h_{n}|^2+\sigma_v^2}\right),\ \ n\in[1:N].
			\label{eq:marginal_E}
		\end{gather}
		\fi
		
		In the special case of an uncorrelated source signal $\mathbf{s}$, in which $\lambda_{m}=\sigma_{s}^2$ for all $m\in[1:M]$, we have:
		\if\mycmd1
		\begin{gather}
			p^1(E_{n}|\mathbf{h})=\left(\frac{\beta(M)}{\sigma_s^2|h_{n}|^2+\sigma_v^2}\right)^M\frac{E_{n}^{M-1}}{(M-1)!}\exp\left(-\frac{\beta(M)E_{n}}{\sigma_s^2|h_{n}|^2+\sigma_v^2}\right),\ \ n\in[1:N],
			\label{eq:marginal_E_white}
		\end{gather}
		\else
		\begin{gather}
			p^1(E_{n}|\mathbf{h})=\left(\frac{\beta(M)}{\sigma_s^2|h_{n}|^2+\sigma_v^2}\right)^M\frac{E_{n}^{M-1}}{(M-1)!}\nonumber\\
			\times\exp\left(-\frac{\beta(M)E_{n}}{\sigma_s^2|h_{n}|^2+\sigma_v^2}\right),\ \ n\in[1:N],
			\label{eq:marginal_E_white}
		\end{gather}
		\fi
		that is, a central chi-square random variable with $2M$ degrees of freedom and mean $\frac{M(\sigma_s^2|h_{n}|^2+\sigma_v^2)}{\beta(M)}$. Although the marginal densities are important, in order to implement a hypothesis test for detecting the source signal, we need the full joint PDF of the energy measurements at the sensors. Although, as we pointed out above, this problem is hard, some insights can be obtained when $M$ is sufficiently large.
		
		\subsection{Approximation of the joint density of $\mathbf{E}$ for the slow fading scenario}
		It is important to analyze, using the characteristic function $\Psi^1(\boldsymbol{\omega}|\mathbf{h})$ in (\ref{eq:CF_uno}), the case in which $\beta(M)$ is large. This will happen when $M$ is large as we have assumed that $\beta(M)$ is an increasing function of $M$. Note that\footnote{Note that $M$ needs to be a positive integer. This is satisfied considering the largest integer smaller or equal to $WT$.} $M\approx WT$, where $W$ is the bandwidth of the continuous time version of the source signal and $T$ is the analog time duration of the sensing window in which the energy measurement in each sensor is done \cite{Landau_Pollak_1962}, \cite{Urkowitz_1967}. For practical applications (e.g., cognitive radio \cite{lunden2015spectrum}, \cite{Liang_Zeng_Peh_Hoang_2008}) the time-bandwidth product $WT$ will not be small, and the following derivations will be useful. Under the assumption of $M$ large, we will use the following first-order approximation of the exponential function:
		\begin{equation}
			1-j\tfrac{\lambda_{m}}{\beta(M)}\sum_{n=1}^N\frac{|h_{n}|^2\omega_n}{1-j\frac{\omega_n\sigma_v^2}{\beta(M)}}\approx \exp\!\left(\!-j\tfrac{\lambda_{m}}{\beta(M)}\sum_{n=1}^N\frac{|h_{n}|^2\omega_n}{1-j\frac{\omega_n\sigma_v^2}{\beta(M)}}\right),
			\label{eq:approx_1}
		\end{equation}
		for $m\in[1:M]$ and $n\in[1:N]$. Under this approximation, $\Psi^1(\boldsymbol{\omega}|\mathbf{h})\approx\hat{\Psi}^1(\boldsymbol{\omega}|\mathbf{h})$ where:
		\begin{equation}
			\hat{\Psi}^1(\boldsymbol{\omega}|\mathbf{h})\equiv \frac{\prod_{m=1}^{M}\exp\left(j\frac{\lambda_{m}}{\beta(M)}\sum_{n=1}^N\frac{|h_{n}|^2\omega_n}{1-j\frac{\omega_n\sigma_v^2}{\beta(M)}}\right)}{\prod_{n=1}^N\left(1-j\frac{\omega_n\sigma_v^2}{\beta(M)}\right)^{M}}.    
			\label{eq:approx_2}
		\end{equation}
		Using that $\sum_{m=1}^{M}\lambda_{m}=\mbox{Tr}\left({\boldsymbol{\Sigma}_{\mathbf{s}}}\right)$ we obtain:
		\begin{equation}
			\hat{\Psi}^1(\boldsymbol{\omega}|\mathbf{h})=\prod_{n=1}^N \frac{\exp\left(\frac{j}{\beta(M)}\frac{\mbox{Tr}\left({\boldsymbol{\Sigma}_{\mathbf{s}}}\right)|h_{n}|^2\omega_n}{1-j\frac{\omega_n\sigma_v^2}{\beta(M)}}\right)}{\left(1-j\frac{\omega_n\sigma_v^2}{\beta(M)}\right)^{M}}.    
			\label{eq:approx_2b}
		\end{equation}

		For a large but fixed value of $M$, notice that the approximation in (\ref{eq:approx_1}) is a very good one when $|\omega_n|$, $n\in[1:N]$ are small. Clearly, for larger values, the approximation is not so good. However, the denominator in (\ref{eq:approx_2b}) increases rapidly  with $|\omega_n|$, $n\in[1:N]$ when $M$ is large and the scaling $\beta(M)$ is chosen carefully. This behaviour seems to have the net effect that  (\ref{eq:approx_2b}) approximates the true CF (\ref{eq:CF_uno}) reasonably well  over large regions in the $\boldsymbol{\omega}$ space. In Section \ref{sec:analysis}, we will rigorously show that this is indeed the case by providing a bound for $\sup_{\mathbf{E}\in\mathbb{R}^N_{\geq 0}}|p^1(\mathbf{E}|\mathbf{h})-\hat{p}^{1}(\mathbf{E}|\mathbf{h})|$ and obtaining its scaling behaviour with $M$.

		The most striking fact about the last expression is that, as $M$ grows, the entries of $\mathbf{E}$ becomes less statistically dependent between them, as (\ref{eq:approx_2b}) is the CF of $N$ independent random variables, each with characteristic function given by:
		\begin{equation}
			\hat{\psi}^1(\omega_n|h_n)\equiv\frac{\exp\left(\frac{j}{\beta(M)}\frac{\mbox{Tr}\left({\boldsymbol{\Sigma}_{\mathbf{s}}}\right)|h_{n}|^2\omega_n}{1-j\frac{\omega_n\sigma_v^2}{\beta(M)}}\right)}{\left(1-j\frac{\omega_n\sigma_v^2}{\beta(M)}\right)^{M}},\  n\in[1:N].
			\label{eq:approx_3}
		\end{equation}
		It is well known that (\ref{eq:approx_3}) is the characteristic function of a non-central chi-square random variable. In more precise terms, anti-transforming (\ref{eq:approx_2b}) we get $p^1(\mathbf{E}|\mathbf{h})\approx \hat{p}^1(\mathbf{E}|\mathbf{h})$, where
		
		%\begin{equation}
		%\label{eq:approx_pdf1}
		%\end{equation}
		
		\begin{align}
			\if\mycmd1
			\hat{p}^1(\mathbf{E}|\mathbf{h})\equiv\prod_{n=1}^N \tfrac{\beta(M)}{\sigma_v^2}\exp\left[-\tfrac{\beta(M)}{\sigma_v^2}\Big(E_{n}+\tfrac{\mbox{Tr}\left({\boldsymbol{\Sigma}_{\mathbf{s}}}\right)|h_n|^2}{\beta(M)}\Big)\right]\!\left(\!\tfrac{\beta(M)E_{n}}{\mbox{Tr}\left({\boldsymbol{\Sigma}_{\mathbf{s}}}\!\right)|h_n|^2}\right)^{\frac{M-1}{2}} \!\! I_{M-1}\!\left(\!\tfrac{2}{\sigma_v^2}\sqrt{\beta(M)\mbox{Tr}\left({\boldsymbol{\Sigma}_{\mathbf{s}}}\right)|h_n|^2E_{n}}\right),
			\else
			& \hat{p}^1(\mathbf{E}|\mathbf{h})\equiv\prod_{n=1}^N \tfrac{\beta(M)}{\sigma_v^2}\exp\left[-\tfrac{\beta(M)}{\sigma_v^2}\Big(E_{n}+\tfrac{\mbox{Tr}\left({\boldsymbol{\Sigma}_{\mathbf{s}}}\right)|h_n|^2}{\beta(M)}\Big)\right]\nonumber \\
			& \times\!\left(\!\tfrac{\beta(M)E_{n}}{\mbox{Tr}\left({\boldsymbol{\Sigma}_{\mathbf{s}}}\!\right)|h_n|^2}\right)^{\frac{M-1}{2}} \!\! I_{M-1}\!\left(\!\tfrac{2}{\sigma_v^2}\sqrt{\beta(M)\mbox{Tr}\left({\boldsymbol{\Sigma}_{\mathbf{s}}}\right)|h_n|^2E_{n}}\right),
			\label{eq:approx_pdf1}
			\fi
		\end{align}
		where $I_{M-1}$ is the modified Bessel function of the first kind and order $M-1$.
		\begin{remark}
			It is clear that under $\mathcal{H}_0$, the components of $\mathbf{E}$ are independent and identically distributed. In addition, it is important to observe that, having $M$ finite but large, allows us to approximately consider that $\{E_{n}\}$ are independent also under $\mathcal{H}_1$. However, (\ref{eq:approx_pdf1}) is different to the product of its marginals, given by (\ref{eq:marginal_E}). This shows that the result of the approximation when $M$ is finite but large is not trivial.
		\end{remark}
		\begin{remark}
			In applications for which $M$ is large and (\ref{eq:approx_1}) leads to a tight CF approximation (\ref{eq:approx_2b}), we can use (\ref{eq:approx_pdf1}) as a substitute of the true (unknown and intractable) joint PDF of $\mathbf{E}$ conditioned to $\mathbf{h}$ and $\mathcal{H}_1$, for building an approximate LRT for the detection problem at hand. This has a twofold advantage. In the first place, although not exact, the approximate likelihood can be easily computed in closed form and we do not have to resort to any series approximation, which actually are not readily available for the considered general scenario. Secondly, as the approximation naturally leads to a factorized PDF, it is well suited for distributed detection scenarios. This is because each node is able to compute its local statistic without cooperation with other nodes, while the final test statistic is the network-wide average of those local statistics, which implies relatively low communication (energy and bandwidth) resources. Notice that likelihood ratio tests using non-product PDF will typically demand many communications resources and are not suitable for distributed scenarios.  
			This will be the case even when, through the use of the multivariate central limit theorem, and selecting $\beta(M)=\frac{1}{M}$, the energy measurements are approximated by a multivariate Gaussian PDF, as it is done in several works  \cite{Maya_2021,Liang_Zeng_Peh_Hoang_2008, Unnikrishnan_Veeravalli_2008,maya2022fading}. In those cases, the likelihood ratio will be a quadratic form depending on the precision matrix (the inverse of the covariance matrix), which will be very costly to compute in distributed scenarios \cite{Wiesel_Hero_2012}.
		\end{remark}
		
		\subsection{Approximation of the joint density of $\mathbf{E}$ in the fast fading scenario}
		In this case, we consider that the channel gains $h_n$ are  random variables. These random values remain fixed during a certain measurement window, but in the next window, they change again according to the same distribution. Contrary to the slow-fading case, here the channel gains are different for each window, and this has to be taken into account. {Using the independence assumption of $\mathbf{E}_l$ across the $L$ measurements windows and that each $\mathbf{E}_l$ conditioned on $\mathbf{h}_l$ is independent of the vector channel gains at the other measurements windows, the approximate joint PDF of the $L$ vector energy measurements $\mathbf{E}_1,\mathbf{E}_2,\dots,\mathbf{E}_L$ conditioned on those channel gains can be expressed as:}
		\begin{equation}
			\label{eq:approx_density_fading}
			\hat{p}^{1}(\mathbf{E}_1,\mathbf{E}_2,\dots,\mathbf{E}_L|\mathbf{h}_1,\mathbf{h}_2,\dots,\mathbf{h}_L)=\prod_{l=1}^{L}\hat{p}^{1}(\mathbf{E}_l|\mathbf{h}_l),
		\end{equation}
		where each $\hat{p}^{1}(\mathbf{E}_l|\mathbf{h}_l)$ for $l\in[1:L]$ is given by (\ref{eq:approx_pdf1}). Clearly, assuming that the $L$ channel gains are known is not a reasonable hypothesis. Assuming that they are unknown, and to devise a test like the GLRT is possible. However, its performance will not be good, as the number of parameters to estimate is $NL$, which is equal to the number of total energy measurements. Another possibility is to average (\ref{eq:approx_density_fading}) over $\mathbf{h}_1,\mathbf{h}_2,\dots, \mathbf{h}_L$ and define\footnote{This averaging is only needed for the energy measurements PDF under $\!\mathcal{H}_1$, due to the resulting PDF under $\mathcal{H}_0$ does not depend on the channel gains.}:
		\begin{equation}
			\label{eq:approx_density_averaged}
			\hat{p}^{1}(\mathbf{E}_1,\mathbf{E}_2,\dots,\mathbf{E}_L)\equiv\mathbb{E}\left[\prod_{l=1}^{L}\hat{p}^{1}(\mathbf{E}_l|\mathbf{h}_l)\right],
		\end{equation}
		where expectation is with respect to $p(\mathbf{h}_1,\mathbf{h}_2,\dots,\mathbf{h}_L)$, the joint PDF of the channel vector gains. Assuming a high-mobility scenario, it is reasonable to consider a fast-fading model \cite{TseViswanath}, where $\mathbf{h}_l$, $l\in[1:L]$ are assumed to be independent and identically distributed random vectors. This allows us to write:
		\begin{equation}
			\label{eq:approx_density_averaged2}
			\hat{p}^{1}(\mathbf{E}_1,\mathbf{E}_2,\dots,\mathbf{E}_L)=\prod_{l=1}^{L}\mathbb{E}\left[\hat{p}^{1}(\mathbf{E}_l|\mathbf{h}_l)\right],
		\end{equation}
		Then, we only need to analyze $\mathbb{E}\left[\hat{p}^{1}(\mathbf{E}_l|\mathbf{h}_l)\right]$ for an arbitrary $l\in[1:L]$.
		We will also assume that nodes are separated well enough. This means that the gains $h_{n}$ (in the following, as we will be again analyzing an arbitrary measurement window, we will drop the sub-index $l$) with $n\in[1:N]$  can be modeled as independent random variables. Under this assumption:
		\begin{equation}
			\mathbb{E}\left[\hat{p}^{1}(\mathbf{E}|\mathbf{h})\right]=\int \hat{p}^{1}(\mathbf{E}|\mathbf{h})\prod_{n=1}^{N}p_n(h_n)dh_1 dh_2\dots dh_N,
			\label{eq:approx_density_averaged3}
		\end{equation}
		where $p_n(h_n)$ is the PDF of the channel gain $h_n$. Several channel distributions of interest can be considered, such as Rayleigh and Nakagami \cite{RappaportWireless}. In the following, we will consider that the channel gains are Rayleigh distributed (non-line-of-sight propagation scenario). Then, the squared gains $|h_n|^2$ are exponentially distributed according to:
		\begin{equation}
			p_n(|h_n|^2)=\frac{1}{\sigma_{n}^2}\exp\left(-\frac{|h_n|^2}{\sigma_n^2}\right),\ \ n\in[1:N],
			\label{eq:Rayleigh}
		\end{equation}
		where $\mathbb{E}\left[|h_n|^2\right]=\sigma_n^2>0$. It is not easy to perform direct integration in (\ref{eq:approx_density_averaged3}). However, in the next lemma, using the CF in (\ref{eq:approx_2b}), we will be able to obtain $\mathbb{E}\left[\hat{p}^{1}(\mathbf{E}|\mathbf{h})\right]$. The proof is relegated to Appendix \ref{ap:fading}.
		\begin{lemma}
			Assuming that $p_n(|h_n|^2)$ for $n\in [1:N]$ is given by (\ref{eq:Rayleigh}), and $\hat{p}^{1}(\mathbf{E}|\mathbf{h})$ is given by (\ref{eq:approx_pdf1}), we have that (\ref{eq:approx_density_averaged3}) can be written as:
			\if\mycmd1
			\begin{gather}
				\mathbb{E}\left[\hat{p}^{1}(\mathbf{E}|\mathbf{h})\right]\!=\!\prod_{n=1}^N\frac{\beta(M)}{\Gamma(M-1)}\frac{\left(\sigma_v^2+\mbox{Tr}\left(\boldsymbol{\Sigma_s}\right)\sigma_n^2 \right)^{M-2}}{\left(\mbox{Tr}\left(\boldsymbol{\Sigma_s}\right)\sigma_n^2\right)^{M-1}}\exp\!\left(\!-\tfrac{\beta(M)E_n}{\sigma_v^2+\mbox{Tr}\left(\boldsymbol{\Sigma_s}\right)\sigma_n^2}\!\right)\!\gamma\left(\!M\!-\!1,\!\tfrac{\beta(M)\mbox{Tr}\left(\boldsymbol{\Sigma_s}\right)\sigma_n^2}{\sigma_v^2\left(\sigma_v^2+\mbox{Tr}\left(\boldsymbol{\Sigma_s}\right)\sigma_n^2\right)}E_n\!\right),
				\label{eq:pdf_fad}
			\end{gather}
			\else
			\begin{gather}
				\mathbb{E}\left[\hat{p}^{1}(\mathbf{E}|\mathbf{h})\right]=\prod_{n=1}^N\frac{\beta(M)}{\Gamma(M-1)}\frac{\left(\sigma_v^2+\mbox{Tr}\left(\boldsymbol{\Sigma_s}\right)\sigma_n^2 \right)^{M-2}}{\left(\mbox{Tr}\left(\boldsymbol{\Sigma_s}\right)\sigma_n^2\right)^{M-1}}\nonumber\\
				\times\exp\!\left(\!-\tfrac{\beta(M)E_n}{\sigma_v^2+\mbox{Tr}\left(\boldsymbol{\Sigma_s}\right)\sigma_n^2}\!\right)\!\gamma\left(\!M\!-\!1,\!\tfrac{\beta(M)\mbox{Tr}\left(\boldsymbol{\Sigma_s}\right)\sigma_n^2}{\sigma_v^2\left(\sigma_v^2+\mbox{Tr}\left(\boldsymbol{\Sigma_s}\right)\sigma_n^2\right)}E_n\!\right),
				\label{eq:pdf_fad}
			\end{gather}
			\fi
			where $\gamma(\alpha,x)\equiv\int_{0}^{x}t^{\alpha-1}e^{-t}dt$, $\alpha,x\geq 0$ is the incomplete Gamma function.
			\label{lemma:fading}
		\end{lemma}
		
		The result of Lemma \ref{lemma:fading}, jointly with (\ref{eq:approx_density_averaged2}), allows us to express an approximation to the true joint PDF  for the full $L$ measurements windows under $\mathcal{H}_1$ in the fast fading scenario. In this way, we can use this approximation to implement a LRT as in (\ref{eq:LR}) also for this scenario, in addition to the slow-fading one. The only issue that remains to be analyzed is the case of the unknown parameters in the derived PDF approximations in a practical application. In the slow fading scenario these parameters are mainly the fixed channel gains $\mathbf{h}$ across the measurement windows.  Meanwhile, in the fast fading scenario, the unknown parameters will  be the values of $\sigma_n^2$, $n\in[1:N]$ which are also fixed for each measurement window. This will be studied in Section \ref{sec:test_approx}.
		
		\section{Theoretical analysis of the approximations}
		\label{sec:analysis}
		
		The approximations (\ref{eq:approx_pdf1}) and (\ref{eq:pdf_fad}) to the joint PDF of the energies sensed at the sensor sites during a measurement window seems to be satisfying from the practical point of view: they are closed-form expressions and are built on the product of PDFs, each of which can be computed locally at each sensor site.
		However, it is important to have some theoretical guarantees for these approximations. This is important on its own, but it is also motivated by the fact that no closed-form or even infinite series expressions are available for the exact joint PDF of $\mathbf{E}$ under $\mathcal{H}_1$. Notice that even in the case where an infinite series expression is available for the mentioned joint PDF, as for a particular case of the considered problem \cite{Morales-Jimenez_Paris_Entrambasaguas_Wong_2011}, the fact that $\mathbf{E}$ is an $N$-dimensional vector, where the amount of sensor nodes $N$ is typically large, poses numerical issues for its computation.
		Moreover, understanding how good these approximations are for a given value of $M$ permits to choose a proper energy scaling $\beta(M)$. Our main interest will be to obtain a meaningful bound for $\sup_{\mathbf{E}\in\mathbb{R}^N_{\geq 0}}|p^1(\mathbf{E}|\mathbf{h})-\hat{p}^{1}(\mathbf{E}|\mathbf{h})|$ as a function of $M$.
		This is loosely connected to the idea of a local limit result \cite{Petrov_1975} for the true density $p^1(\mathbf{E}|\mathbf{h})$ when $M\rightarrow\infty$. However, some differences are worth noting. In the first place, we are not interested in the limit of $p^1(\mathbf{E}|\mathbf{h})$ or $\hat{p}^{1}(\mathbf{E}|\mathbf{h})$ when $M\rightarrow\infty$ (which could not even be well defined). We are only interested in computing how well these two expressions match when $M$ is large. That is, we want some estimate on the rate at which these two expressions get closer. In the second place, under $\mathcal{H}_1$, and for each $n\in[1:N]$, $E_{n}\equiv\frac{1}{\beta(M)}\|\mathbf{y}_{n}\|^2$ is not the sum of independent random variables as the components of each $\mathbf{y}_n$ are dependent. Most results available in the literature about local limit theorems, or even limit theorems for distribution functions (like the celebrated Berry-Essen Theorem \cite{Feller_1971}) are restricted to the independent and identically distributed case. It should be clear that obtaining meaningful bounds on $\sup_{\mathbf{E}\in\mathbb{R}^N_{\geq 0}}|p^1(\mathbf{E}|\mathbf{h})-\hat{p}^{1}(\mathbf{E}|\mathbf{h})|$ could be more difficult or would require stronger technical conditions that getting similar results but for the distribution functions. However, optimal LRTs depend on the PDFs under both hypotheses, and not on the distribution functions. For this reason, we will study the term $\sup_{\mathbf{E}\in\mathbb{R}^N_{\geq 0}}|p^1(\mathbf{E}|\mathbf{h})-\hat{p}^{1}(\mathbf{E}|\mathbf{h})|$. In our case, we will exploit the structure of the  CFs given by (\ref{eq:CF_uno}) and (\ref{eq:approx_2b}).
		
		From the fact that a PDF can be written as the $N$-dimensional inverse Fourier transform of its CF, we can easily get for all $\mathbf{E}\in\mathbb{R}^N_{\geq 0}$:
		\begin{equation}
			\big|p^1(\mathbf{E}|\mathbf{h})\!-\hat{p}^{1}(\mathbf{E}|\mathbf{h})\big|\leq \!\tfrac{1}{(2\pi)^N}\int_{\mathbb{R}^N}\!\big|\Psi^1(\boldsymbol{\omega}|\mathbf{h})\!-\!\hat{\Psi}^1(\boldsymbol{\omega}|\mathbf{h})\big|d\boldsymbol{\omega},
			\label{eq:CF_bound_max}
		\end{equation}
		from which we conclude that the $L_1$ distance between the CFs $\Psi^1(\boldsymbol{\omega}|\mathbf{h})$ and $\hat{\Psi}^1(\boldsymbol{\omega}|\mathbf{h})$ is an upper bound for $\sup_{\mathbf{E}\in\mathbb{R}^N_{\geq 0}}|p^1(\mathbf{E}|\mathbf{h})-\hat{p}^{1}(\mathbf{E}|\mathbf{h})|$. Let us define:
		\begin{equation}
			z_m(\boldsymbol{\omega})\equiv j\frac{\lambda_m}{\beta(M)}\sum_{n=1}^N\frac{|h_n|^2\omega_n}{1-j\frac{\sigma_v^2\omega_n}{\beta(M)}},\ m\in[1:M].
			\label{eq:z_def}
		\end{equation}
		Using  (\ref{eq:CF_uno}) and (\ref{eq:approx_2b}), we can write:
		\if\mycmd1
		\begin{equation}
			\Big|\Psi^1(\boldsymbol{\omega}|\mathbf{h})-\hat{\Psi}^1(\boldsymbol{\omega}|\mathbf{h})\Big|\leq
			\prod_{n=1}^N \Big|1-j\frac{\sigma_v^2\omega_n}{\beta(M)}\Big|^{-M}\Big|\prod_{m=1}^M\frac{1}{1-z_m(\boldsymbol{\omega})}-\prod_{m=1}^Me^{z_m(\boldsymbol{\omega})}\Big|.
			\label{eq:L1_CF1}
		\end{equation}
		\else
		\begin{multline}
			\Big|\Psi^1(\boldsymbol{\omega}|\mathbf{h})-\hat{\Psi}^1(\boldsymbol{\omega}|\mathbf{h})\Big|\leq
			\prod_{n=1}^N \Big|1-j\frac{\sigma_v^2\omega_n}{\beta(M)}\Big|^{-M}\\
			\times\Big|\prod_{m=1}^M\frac{1}{1-z_m(\boldsymbol{\omega})}-\prod_{m=1}^Me^{z_m(\boldsymbol{\omega})}\Big|.
			\label{eq:L1_CF1}
		\end{multline}
		\fi
		The following lemma is a consequence of the fact that $\mathfrak{Re}(z_m(\boldsymbol{\omega}))\leq 0$ for all $\boldsymbol{\omega}\in\mathbb{R}^N$ and $m\in[1:M]$. %Its proof is not difficult and for that reason it is omitted.
		and it is proved in Appendix \ref{ap:z}:
		\begin{lemma}
			\label{lemma:bound_z}
			Let $z_m(\boldsymbol{\omega})$, $m\in[1:M]$ defined in (\ref{eq:z_def}). Then for all $\boldsymbol{\omega}\in\mathbb{R}^N$:
			\if\mycmd1
			\begin{equation}
				\left|\prod_{m=1}^M\frac{1}{1-z_m(\boldsymbol{\omega})}-\prod_{m=1}^Me^{z_m(\boldsymbol{\omega})}\right|
				\leq \sum_{m=1}^M\min\left\{2,\left|1-z_m(\boldsymbol{\omega})-e^{-z_m(\boldsymbol{\omega})}\right|    \right\}
				\label{eq:lemma_z}
			\end{equation}    
			\else
			\begin{multline}
				\left|\prod_{m=1}^M\frac{1}{1-z_m(\boldsymbol{\omega})}-\prod_{m=1}^Me^{z_m(\boldsymbol{\omega})}\right|\\
				\leq \sum_{m=1}^M\min\left\{2,\left|1-z_m(\boldsymbol{\omega})-e^{-z_m(\boldsymbol{\omega})}\right|    \right\}
				\label{eq:lemma_z}
			\end{multline}   	 
			\fi
		\end{lemma}
		The following lemma is important because it allows us to quantify the error on the first-order Taylor expansion of $e^{-z}$ given by $1-z$ when $\mathfrak{Re}(z)\leq 0$. We want to emphasize that as we are considering a Taylor expansion of an analytic complex valued function special care has to be taken to estimate the remainder of the expansion. The fact that the value $z$ where we are computing the approximation is in a specific region of the complex plane allows us to obtain better estimates of the remainder. The proof can be found in Appendix \ref{ap:exp}.
		
		%As the proof is very simple,  it is also omitted for lack of space. %The proof is relegated to Appendix \ref{ap:exp}.
		\begin{lemma}
			\label{lemma:exp}
			Consider $z\in\mathbb{C}$ such that $\mathfrak{Re}(z)\leq 0$. Then:
			\begin{equation}
				|1-z-e^{-z}|\leq|z|^2e^{-\mathfrak{Re}(z)}
				\label{eq:bound_exp}
			\end{equation}   	 
		\end{lemma}
		At this point we can combine the results from Lemma \ref{lemma:bound_z} and \ref{lemma:exp} to get:
		\if\mycmd1
		\begin{equation}
			\left|\prod_{m=1}^M\frac{1}{1-z_m(\boldsymbol{\omega})}-\prod_{m=1}^Me^{z_m(\boldsymbol{\omega})}\right|
			\leq \sum_{m=1}^M\min\left\{2,|z_m(\boldsymbol{\omega})|^2e^{-\mathfrak{Re}(z_m(\boldsymbol{\omega}))}\right\}.
			\label{eq:comb_lemmas}
		\end{equation}
		\else
		\begin{multline}
			\left|\prod_{m=1}^M\frac{1}{1-z_m(\boldsymbol{\omega})}-\prod_{m=1}^Me^{z_m(\boldsymbol{\omega})}\right|\\
			\leq \sum_{m=1}^M\min\left\{2,|z_m(\boldsymbol{\omega})|^2e^{-\mathfrak{Re}(z_m(\boldsymbol{\omega}))}\right\}.
			\label{eq:comb_lemmas}
		\end{multline}
		\fi
		It is easy to show that $\max_{m\in[1:M]}\left(-\inf_{\boldsymbol{\omega}\in\mathbb{R}^N}\mathfrak{Re}(z_m(\boldsymbol{\omega}))\right)=\frac{\lambda_{\rm{max}}}{\sigma_v^2}\sum_{n=1}^N|h_n|^2$ where $\lambda_{\rm{max}}$ is the maximum eigenvalue of $\boldsymbol{\Sigma_s}$. Then, from (\ref{eq:comb_lemmas}) we can get:
		\if\mycmd1
		\begin{equation}
			\left|\prod_{m=1}^M\frac{1}{1-z_m(\boldsymbol{\omega})}-\prod_{m=1}^Me^{z_m(\boldsymbol{\omega})}\right|
			\leq\min\left\{2M, e^{\frac{\lambda_{\rm{max}}}{\sigma_v^2}\sum_{n=1}^N|h_n|^2}\sum_{m=1}^M|z_m(\boldsymbol{\omega})|^2\right\}.
			\label{eq:bound_z_aux_2}
		\end{equation}
		\else
		\begin{multline}
			\left|\prod_{m=1}^M\frac{1}{1-z_m(\boldsymbol{\omega})}-\prod_{m=1}^Me^{z_m(\boldsymbol{\omega})}\right|\\
			\leq\min\left\{2M, e^{\frac{\lambda_{\rm{max}}}{\sigma_v^2}\sum_{n=1}^N|h_n|^2}\sum_{m=1}^M|z_m(\boldsymbol{\omega})|^2\right\}.
			\label{eq:bound_z_aux_2}
		\end{multline}
		\fi
		
		Using (\ref{eq:z_def}), the term $\sum_{m=1}^M|z_m(\boldsymbol{\omega})|^2$ can be bounded as:
		\if\mycmd1
		\begin{equation}
			\sum_{m=1}^M|z_m(\boldsymbol{\omega})|^2=\sum_{m=1}^M\left|j\frac{\lambda_m}{\beta(M)}\sum_{n=1}^N\frac{|h_n|^2\omega_n}{1-j\frac{\sigma_v^2\omega_n}{\beta(M)}}\right|^2\leq\frac{\delta(\mathbf{h})M}{\beta^2(M)}\!\sum_{p=1}^N \!\sum_{r=1}^N \!\frac{|\omega_p||\omega_r|}{\sqrt{1\!+\!\frac{\sigma_v^4\omega_p^2}{\beta^2(M)}}\sqrt{1\!+\!\frac{\sigma_v^4\omega_r^2}{\beta^2(M)}}},\nonumber
		\end{equation}
		\else
		\begin{eqnarray}
			\sum_{m=1}^M|z_m(\boldsymbol{\omega})|^2\!\!\!\!\!&=&\!\!\!\!\!\!\sum_{m=1}^M\left|j\frac{\lambda_m}{\beta(M)}\sum_{n=1}^N\frac{|h_n|^2\omega_n}{1-j\frac{\sigma_v^2\omega_n}{\beta(M)}}\right|^2\nonumber\\
			&\leq&\!\!\!\tfrac{\delta(\mathbf{h})M}{\beta^2(M)}\!\sum_{p=1}^N \!\sum_{r=1}^N \!\frac{|\omega_p||\omega_r|}{\sqrt{1\!+\!\frac{\sigma_v^4\omega_p^2}{\beta^2(M)}}\sqrt{1\!+\!\frac{\sigma_v^4\omega_r^2}{\beta^2(M)}}},\nonumber
		\end{eqnarray}
		\fi
		where $\delta(\mathbf{h})\equiv\lambda_{\rm max}^2\max_{n\in[1:N]}|h_n|^4$. In this way we can bound (\ref{eq:bound_z_aux_2}) as:
		\if\mycmd1
		\begin{equation}
			\left|\prod_{m=1}^M\frac{1}{\left(1-z_m(\boldsymbol{\omega})\right)}-\prod_{m=1}^Me^{z_m(\boldsymbol{\omega})}\right|
			\leq\min\left\{2M, \tfrac{\delta'(\mathbf{h})M}{\beta^2(M)}\sum_{p=1}^N \!\sum_{r=1}^N \!\frac{|\omega_p||\omega_r|}{\sqrt{1\!+\!\frac{\sigma_v^4\omega_p^2}{\beta^2(M)}}\sqrt{1\!+\!\frac{\sigma_v^4\omega_r^2}{\beta^2(M)}}}\right\},
			\label{eq:bound_z_aux_3}
		\end{equation}
		\else
		\begin{multline}
			\left|\prod_{m=1}^M\frac{1}{\left(1-z_m(\boldsymbol{\omega})\right)}-\prod_{m=1}^Me^{z_m(\boldsymbol{\omega})}\right|\\
			\leq\min\left\{2M, \tfrac{\delta'(\mathbf{h})M}{\beta^2(M)}\sum_{p=1}^N \!\sum_{r=1}^N \!\frac{|\omega_p||\omega_r|}{\sqrt{1\!+\!\frac{\sigma_v^4\omega_p^2}{\beta^2(M)}}\sqrt{1\!+\!\frac{\sigma_v^4\omega_r^2}{\beta^2(M)}}}\right\},
			\label{eq:bound_z_aux_3}
		\end{multline}
		\fi
		where $\delta'(\mathbf{h})\equiv\delta(\mathbf{h})e^{\frac{\lambda_{\rm{max}}}{\sigma_v^2}\sum_{n=1}^N|h_n|^2}$. At this point we can proceed integrating (\ref{eq:L1_CF1}) using the bound in (\ref{eq:bound_z_aux_3}). The next theorem, whose proof is in Appendix \ref{ap:asymp}, follows:
		\begin{theorem}
			\label{theo:asymp}
			The maximal error between $p^1(\mathbf{E}|\mathbf{h})$ and $\hat{p}^{1}(\mathbf{E}|\mathbf{h})$ can be bounded as:
			\if\mycmd1
			\begin{equation}
				\label{eq:finalboundtheo}
				\sup_{\mathbf{E}\in\mathbb{R}^N_{\geq 0}}|p^1(\mathbf{E}|\mathbf{h})-\hat{p}^{1}(\mathbf{E}|\mathbf{h})|=
				\mathcal{O}\left(
				\min\left\{\tfrac{\beta^{N}(M)}{M^{\frac{N}{2}-1}},\delta'(\mathbf{h})\tfrac{\beta^{N}(M)}{M^{\frac{N}{2}}}\right\}\right).
			\end{equation}
			\else
			\begin{multline}
				\label{eq:finalboundtheo}
				\sup_{\mathbf{E}\in\mathbb{R}^N_{\geq 0}}|p^1(\mathbf{E}|\mathbf{h})-\hat{p}^{1}(\mathbf{E}|\mathbf{h})|=\\
				\mathcal{O}\left(
				\min\left\{\tfrac{\beta^{N}(M)}{M^{\frac{N}{2}-1}},\delta'(\mathbf{h})\tfrac{\beta^{N}(M)}{M^{\frac{N}{2}}}\right\}\right).
			\end{multline}
			\fi
		\end{theorem}
		From the above theorem we can obtain the following conclusions for the two scenarios we are considering:
		\subsubsection{Slow fading scenario} When $M$ sufficiently large the second term in the RHS in (\ref{eq:finalboundtheo}) is the tighter term. We see that for $\sup_{\mathbf{E}\in\mathbb{R}^N_{\geq 0}}|p^1(\mathbf{E}|\mathbf{h})-\hat{p}^{1}(\mathbf{E}|\mathbf{h})|\rightarrow 0$ when $M\rightarrow\infty$ for every value of $\mathbf{h}$, we need  $\frac{\beta(M)}{M^{\frac{1}{2}}}\rightarrow 0$. This means that $\beta(M)=M^{1/2-\epsilon}$ for $0<\epsilon<1/2$ would be a valid choice. With this choice we get:
		\begin{equation}
			\sup_{\mathbf{E}\in\mathbb{R}^N_{\geq 0}}|p^1(\mathbf{E}|\mathbf{h})-\hat{p}^{1}(\mathbf{E}|\mathbf{h})|= \delta'(\mathbf{h})\mathcal{O}\left(\frac{1}{M^{N\epsilon}}\right) \label{eq:bound_ff}
		\end{equation}
		which means that the approximation is exponentially fast in $N$, the size of the network.
		\subsubsection{Fast fading scenario} In this case we should proceed with more care. We need to consider a bound on $\sup_{\mathbf{E}\in\mathbb{R}^N_{\geq 0}}|\mathbb{E}\left[p^1(\mathbf{E}|\mathbf{h})\right]-\mathbb{E}\left[\hat{p}^{1}(\mathbf{E}|\mathbf{h})\right]|$. We can write:
		\if\mycmd1
		\begin{equation}
			\label{eq:boundsupin}
			\sup_{\mathbf{E}\in\mathbb{R}^N_{\geq 0}}|\mathbb{E}\left[p^1(\mathbf{E}|\mathbf{h})\right]-\mathbb{E}\left[\hat{p}^{1}(\mathbf{E}|\mathbf{h})\right]|
			\leq\mathbb{E}\left[\sup_{\mathbf{E}\in\mathbb{R}^N_{\geq 0}}|p^1(\mathbf{E}|\mathbf{h})-\hat{p}^{1}(\mathbf{E}|\mathbf{h})|\right].
		\end{equation}
		\else
		\begin{multline}
			\label{eq:boundsupin}
			\sup_{\mathbf{E}\in\mathbb{R}^N_{\geq 0}}|\mathbb{E}\left[p^1(\mathbf{E}|\mathbf{h})\right]-\mathbb{E}\left[\hat{p}^{1}(\mathbf{E}|\mathbf{h})\right]|\\
			\leq\mathbb{E}\left[\sup_{\mathbf{E}\in\mathbb{R}^N_{\geq 0}}|p^1(\mathbf{E}|\mathbf{h})-\hat{p}^{1}(\mathbf{E}|\mathbf{h})|\right].
		\end{multline}
		\fi
		At this point we can use (\ref{eq:finalboundtheo}). As the second term in the RHS of (\ref{eq:finalboundtheo}) includes the term $e^{\frac{\lambda_{\rm{max}}}{\sigma_v^2}\sum_{n=1}^N|h_n|^2}$, the simplest way to guarantee a non-trivial bound, independent of the channel gains distribution, is to use the first term in the RHS of (\ref{eq:finalboundtheo}). We obtain:
		\begin{equation}
			\label{eq:fastbound}
			\sup_{\mathbf{E}\in\mathbb{R}^N_{\geq 0}}|\mathbb{E}\left[p^1(\mathbf{E}|\mathbf{h})\right]-\mathbb{E}\left[\hat{p}^{1}(\mathbf{E}|\mathbf{h})\right]|=\mathcal{O}\left(\frac{\beta^{N}(M)}{M^{\frac{N}{2}-1}}\right).
		\end{equation}
		Clearly choosing $\beta(M)=M^{1/2-\epsilon}$ with $\frac{1}{N}<\epsilon<1/2-\frac{1}{N}$ we have:
		\begin{equation}
			\label{eq:fastbound2}
			\sup_{\mathbf{E}\in\mathbb{R}^N_{\geq 0}}|\mathbb{E}\left[p^1(\mathbf{E}|\mathbf{h})\right]-\mathbb{E}\left[\hat{p}^{1}(\mathbf{E}|\mathbf{h})\right]|=\mathcal{O}\left(\frac{1}{M^{N\left(\epsilon-\frac{1}{N}\right)}}\right).
		\end{equation}
		\begin{remark}
			We see that in the fast fading scenario there is a small penalty in the possible rate of convergence of the maximal error of the approximation. This penalty decrease with the network size. It is worth to mention that this penalty appears because of the way we use  (\ref{eq:finalboundtheo}) in (\ref{eq:boundsupin}). A more thoughtful use of (\ref{eq:finalboundtheo})  or even a direct treatment of $\sup_{\mathbf{E}\in\mathbb{R}^N_{\geq 0}}|\mathbb{E}\left[p^1(\mathbf{E}|\mathbf{h})\right]-\mathbb{E}\left[\hat{p}^{1}(\mathbf{E}|\mathbf{h})\right]|$ using similar arguments to those in Appendix \ref{ap:asymp} will surely lead to a better scaling and even one without any penalty with respect to the slow fading case. However, the result obtained is already sufficient for us and an improvement of this result is out of scope for the present paper.
		\end{remark}
		
		\section{Proposed test statistics}
		\label{sec:test_approx}
		In this section, we define two test statistics for the slow and fast fading scenarios, using the approximations to the joint PDF presented in Sec. III-B and Sec. III-C, respectively. Those expressions depend on parameters that are typically unknown for the setup at hand, as for example, the power of the transmitting source, the channel gains in the slow fading case and the channel variance in the fast fading case\footnote{The value of the noise variance $\sigma_v^2$ is also usually not known. However, it can be easily estimated under source silence periods to obtain a constant false alarm rate test. For this reason, it will be assumed to be known.}. This fact prevents the implementation of the LRT. In addition, the uniformly most powerful test (UMPT) \cite{Levy_Det} typically does not exist for problems with multidimensional unknown parameters as in the case of this work. Therefore, we resort to the GLRT \cite{Levy_Det}, where the unknown parameters are estimated using the maximum likelihood estimator.    
		We will consider the special case in which $\boldsymbol{\Sigma}_{\mathbf{s},l} \equiv \boldsymbol{\Sigma}_{\mathbf{s}}$ for every $l=1,\dots,L$. That is, during the $L$ measurement windows the statistical properties of the source signal are invariant.
		
		\subsection{Approximate GLRT for the slow fading scenario}
		In the slow fading case, we assume that the channel \emph{gains} between the source position and the sensors are invariant during the $L$ measurement windows. Therefore, we have that $h_{n,l}= h_{n}$ for every $n=1,\dots, N$ and $l=1,\dots,L$. We define $c_n\equiv\mbox{Tr}(\boldsymbol{\Sigma}_\mathbf{s})|h_n|^2$ for $n=1,\dots, N$ leading to the test:
		\begin{equation}
			\tilde{T}_\text{GLRT-SF}(\mathbf{E}_1,\dots,\mathbf{E}_L)=\sum_{n=1}^N\sum_{l=1}^L\log \left(\frac{\tilde{p}_{\rm SF}^1(E_{n,l};\hat{c}_n)}{p^0(E_{n,l})}\right) \underset{\mathcal{H}_0}{ \overset{\mathcal{H}_1}{\gtrless}} \tau,
			\label{eq:GLRT-SF}
		\end{equation}
		where, from (\ref{eq:approx_pdf1}), we define:
		\if\mycmd1
		\begin{equation}
			\tilde{p}_{\rm SF}^1(E_{n,l};c_n) \equiv \frac{\beta(M)}{\sigma_v^2}\exp\left[-\frac{\beta(M)}{\sigma_v^2}\Big(E_{n}+\frac{c_n}{\beta(M)}\Big)\right]\left(\!\frac{\beta(M)E_{n}}{c_n}\right)^{\frac{M-1}{2}} \!\! I_{M-1}\!\left(\!\frac{2}{\sigma_v^2}\sqrt{\beta(M)c_n E_{n}}\right)
		\end{equation}
		\else
		\begin{multline}
			\tilde{p}_{\rm SF}^1(E_{n,l};c_n) \equiv \frac{\beta(M)}{\sigma_v^2}\exp\left[-\frac{\beta(M)}{\sigma_v^2}\Big(E_{n}+\frac{c_n}{\beta(M)}\Big)\right]\\
			\!\times\!\left(\!\frac{\beta(M)E_{n}}{c_n}\right)^{\frac{M-1}{2}} \!\! I_{M-1}\!\left(\!\frac{2}{\sigma_v^2}\sqrt{\beta(M)c_n E_{n}}\right)
			\label{eq:aprox_SF}
		\end{multline}
		\fi
		and $\hat{c}_n \equiv \arg\max_{c_n\geq 0} \sum_{l=1}^L\log\tilde{p}_{\rm SF}^1(E_{n,l};c_n)$. We used the fact that the unknown parameters are positive and only affect the energy measurements under $\mathcal{H}_1$. The PDF $p^0(E_{n,l})$ is given by:
		\begin{equation}
			p^0(E_{n,l})\equiv \left(\frac{\beta(M)}{\sigma_v^2}\right)^{\!M}\frac{E_{n}^{M-1}}{(M-1)!}\exp\left(-\frac{\beta(M)E_{n}}{\sigma_v^2}\right).\label{eq:p0nl}
		\end{equation}
		{The threshold $\tau$ is given by the desired probability of false alarm  $P_{\text{fa}}$ depending only on the distribution of the test under $\mathcal{H}_0$ (and independent of the unknown parameters under $\mathcal{H}_1$) and its computed by Monte Carlo simulations using (\ref{eq:p0nl}).}
		
		\subsection{Approximate GLRT for the fast fading scenario}
		In the fast fading case, we assume that the channel \emph{statistics} between the source position and the sensors are invariant during the $L$ measurement windows. Therefore, we have that $\sigma^2_n$ is constant during the whole sensing time. We define $d_n\equiv\mbox{Tr}(\boldsymbol{\Sigma}_\mathbf{s})\sigma^2_n$ for $n=1,\dots, N$ leading to the test:
		\begin{equation}
			\tilde{T}_\text{GLRT-FF}(\mathbf{E}_1,\dots,\mathbf{E}_L)=\sum_{n=1}^N\sum_{l=1}^L\log \left(\frac{\tilde{p}_{\rm FF}^1(E_{n,l};\hat{d}_n)}{p^0(E_{n,l})}\right) \underset{\mathcal{H}_0}{ \overset{\mathcal{H}_1}{\gtrless}} \tau,
			\label{eq:GLRT-FF}
		\end{equation}
		where, from (\ref{eq:pdf_fad}), we define
		\if\mycmd1
		\begin{equation}
			\tilde{p}_{\rm FF}^1(E_{n,l};d_n) \equiv
			\frac{\beta(M)}{\Gamma(M-1)}\frac{\left(\sigma_v^2+d_n \right)^{M-2}} {d_n^{M-1}}
			\exp\!\left(\!-\frac{\beta(M)E_n}{\sigma_v^2+d_n}\!\right)\!\gamma\left(\!M\!-\!1,\!\frac{\beta(M)d_n}{\sigma_v^2\left(\sigma_v^2+d_n\right)}E_n\!\right)
		\end{equation}
		\else
		\begin{multline}
			\tilde{p}_{\rm FF}^1(E_{n,l};d_n) \equiv
			\frac{\beta(M)}{\Gamma(M-1)}\frac{\left(\sigma_v^2+d_n \right)^{M-2}} {d_n^{M-1}}\times\nonumber\\
			\exp\!\left(\!-\frac{\beta(M)E_n}{\sigma_v^2+d_n}\!\right)\!\gamma\left(\!M\!-\!1,\!\frac{\beta(M)d_n}{\sigma_v^2\left(\sigma_v^2+d_n\right)}E_n\!\right)
		\end{multline}
		\fi
		and $\hat{d}_n \equiv \arg\max_{d_n\geq 0} \sum_{l=1}^L\log\tilde{p}_{\rm FF}^1(E_{n,l};d_n)$. {As in the previous case, we used the fact that the unknown parameters are positive and only affect the energy measurements under $\mathcal{H}_1$. The PDF $p^0(E_{n,l})$ is defined in (\ref{eq:p0nl}). Again, the threshold $\tau$ is determined numerically by the desired $P_\text{fa}$ using (\ref{eq:p0nl}).}
		
		\begin{remark}
			
			Both statistics $\tilde{T}_\text{GLRT-FF}(\mathbf{E}_1,\dots,\mathbf{E}_L)$ and $\tilde{T}_\text{GLRT-SF}(\mathbf{E}_1,\dots,\mathbf{E}_L)$ need to be implemented by the network in order to make a decision about the presence or not of the source signal.
			In any of the two network architectures (with or without FC) shown in Fig. \ref{fig:network}, the sensor nodes need reporting communication channels for building the statistics. In both cases, the sensor nodes compute the corresponding inner sum over index $l$ in (\ref{eq:GLRT-SF}) and (\ref{eq:GLRT-FF}). Then, in the case of a network with a FC, the nodes communicate those quantities to the FC, where the outer sum over $n$ in (\ref{eq:GLRT-SF}) and (\ref{eq:GLRT-FF}) is computed. In the case of a network without a FC, an average consensus algorithm (e.g., see \cite{xiao2004fast}, \cite{Maya_2021}) can be used to compute the outer sum cooperatively among the nodes via messages exchanges between the nodes. Once the statistics are computed, the decision is made.
			In this work, given that our main focus is on how to obtain a closed-form approximation of the joint density, and its use to compute the Neyman-Person test, we assume error-free communication channels\footnote{We also do not consider the quantization procedure needed in digital communication schemes.} in each of the network architectures. Channel impairments, as considered for example in \cite{Li_Li_Varshney_2020}, \cite{Viswanathan_Varshney_1997}, \cite{Patel_Ram_Jagannatham_Varshney_2018}, \cite{Chawla_Patel_Jagannatham_Varshney_2019},  are left for future works. More importantly, the approximate factorization of the joint PDF proposed in (\ref{eq:approx_pdf1}) and (\ref{eq:pdf_fad}) allows us to also factorize the LRT in (\ref{eq:LR}) through the spatial index $n$, and efficiently compute the statistics in a distributed scenarios (with or without FC), as explained before, using communication resources only for computing the outer sum over $n$.
	\end{remark}
	
	\section{Numerical results}
	\label{sec:numerical_results}
\begin{figure}[htb!]
		\centering
		\subfigure[][Slow fading.]{\includegraphics[width=\linewidth]{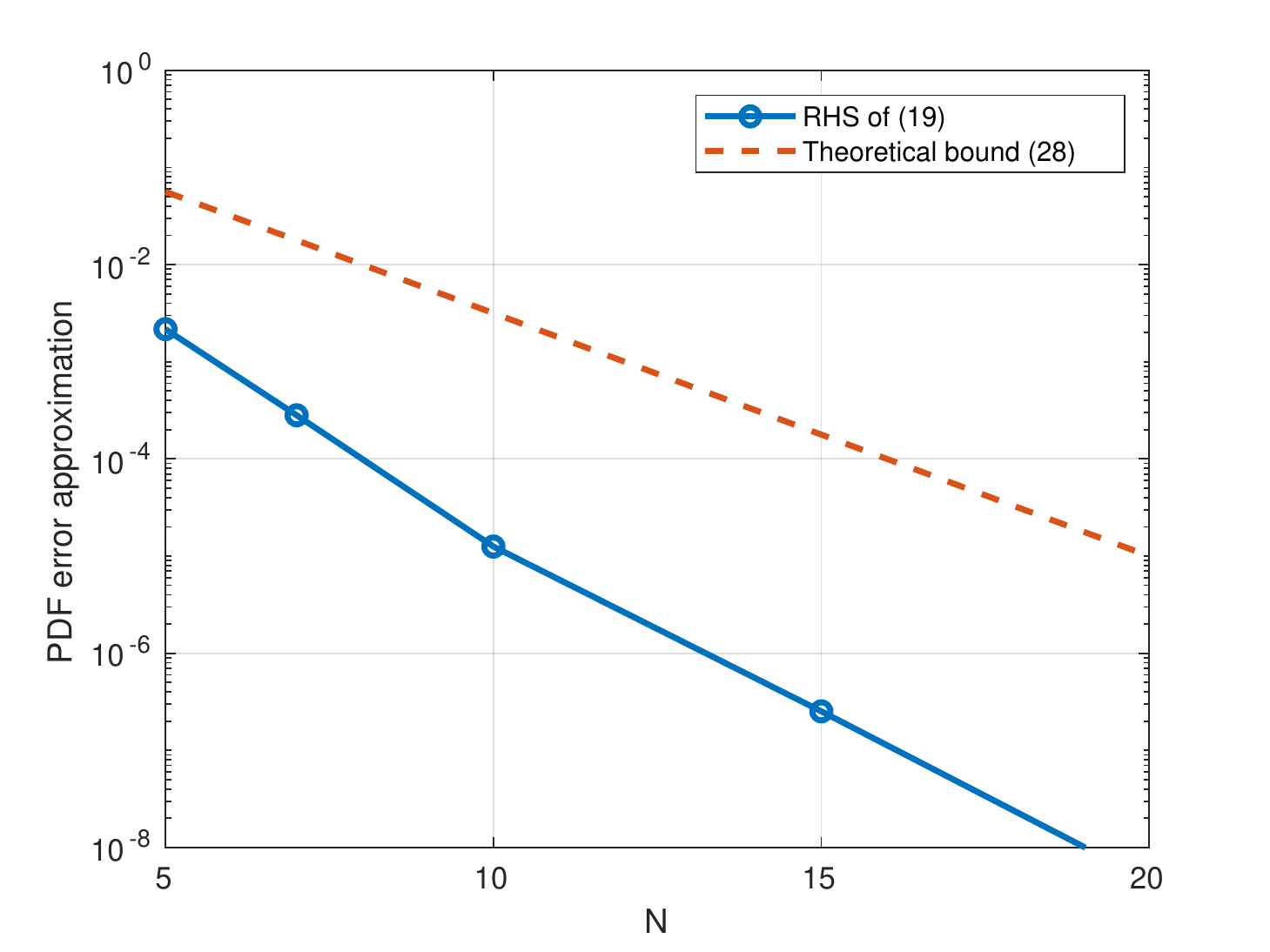}}
		\subfigure[][Fast fading.]{\includegraphics[width=\linewidth]{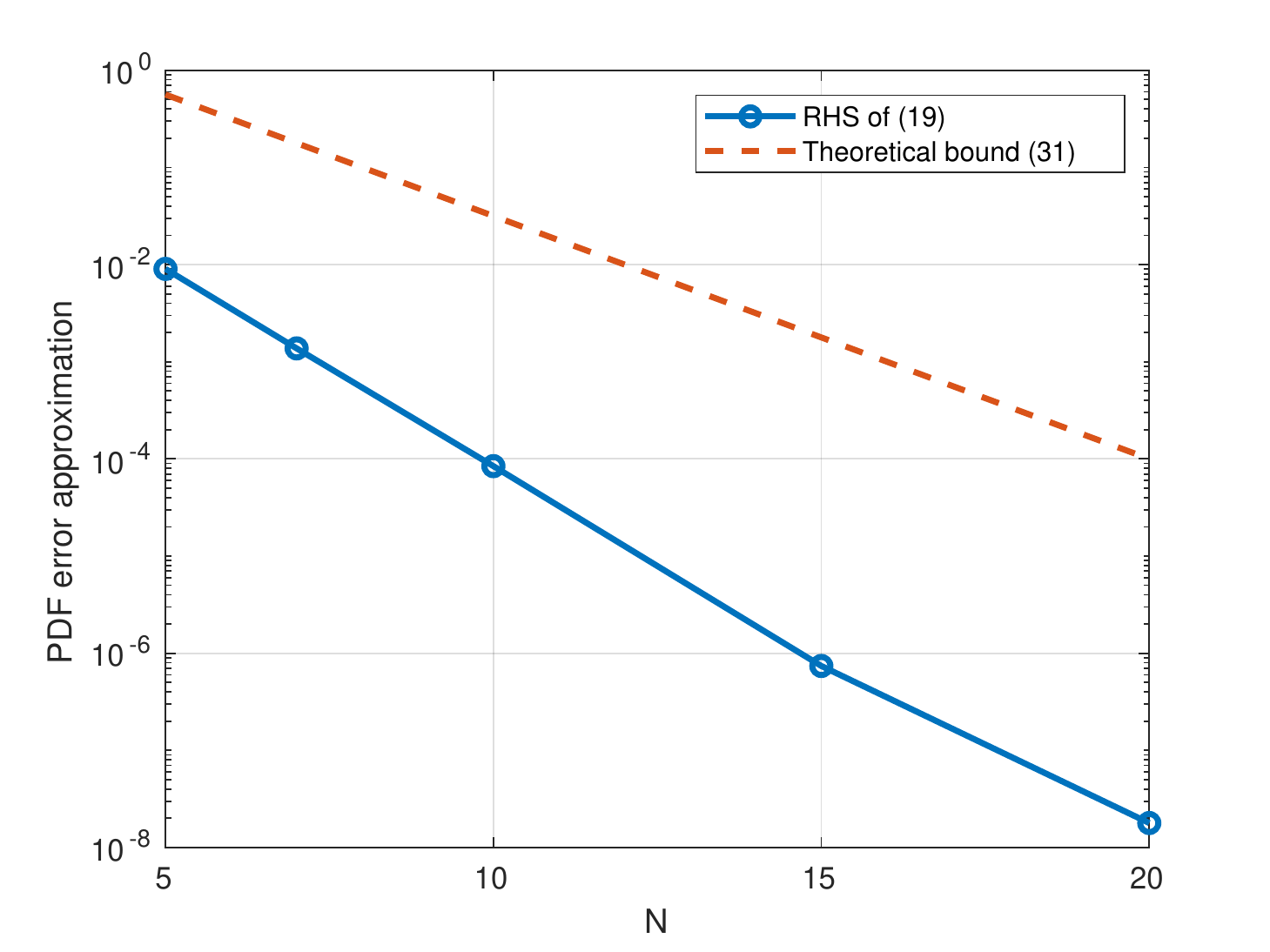}}
		\caption{Evaluation of the joint PDF approximation error bounds for slow and fast fading.}
		\label{fig:app_err}
	\end{figure}

	{In this section, we first assess the theoretical bounds presented in Section IV, and then compute the performance of the algorithms GLRT-SF and GLRT-FF proposed in the previous section.
		\subsection{Theoretical bounds assessment}
		Next, we numerically evaluate the theoretical bounds for the error approximation of the joint PDF of the energy measurements given in (\ref{eq:bound_ff}) and (\ref{eq:fastbound2}). Given that the true joint PDF of the energy measurements are unknown for the two considered scenarios (fast and slow fading), we numerically compute the bound in (\ref{eq:CF_bound_max}) based on the true characteristic functions of the corresponding joint PDFs. The right hand side (RHS) of (\ref{eq:CF_bound_max}), which involves a multidimensional integral, is computed via Monte Carlo simulations with $10^7$ trials. On the other hand, as the theoretical bounds (\ref{eq:bound_ff}) and (\ref{eq:fastbound2}) are expressed in terms of the big-O notation, they do not provide the positive proportional constant of the bounds. Therefore, we are here interested in comparing the slopes of the curves. The bounds are plotted in Fig. \ref{fig:app_err} in semi-log graphs. The scenario parameters are selected as in the following section for the Gaussian source, with $\text{SNR}=10$ dB and $\epsilon=\frac{1}{4}$ (see also Table \ref{tab:param} for other relevant parameters). These curves verify that the error approximation of the joint PDF of the energy measurements decays exponentially with the number of nodes $N$. The slightly difference in the slopes in each scenario can be explained in terms of the bounding techniques used in the derivation of the bounds, which are not tight. However, it is clear that the approximation proposed is very good for reasonable values of $M$ and $N$.
		% Tighter bounds are out of the scope of the present work.
	}
	
	\subsection{Performance analysis of GLRT-SF and GLRT-FF}
	% La info de este parrafo esta en el remark de arriba.
	%We will assume that the sensor network is able to build both statistics (\ref{eq:GLRT-SF}) and (\ref{eq:GLRT-FF}) in a fully-distributed fashion using some consensus-based algorithm, as for example \cite{xiao2004fast} (see also \cite{Maya_2021}), to compute the outer sum over the index $n$. The inner sum over the index $l$ and the estimation of the corresponding parameter can be done locally at each node, without exchanging information with other nodes.
	We consider the problem presented in Sec. II for both slow fading and fast fading scenarios. {This problem can be motivated by a cooperative spectrum sensing application, where a set of secondary nodes senses a frequency band reserved to a primary user. The main idea is that, if the primary user is not using the band, then the secondary users could use it. As the primary user has priority in using this band, it is very important to have a robust detection procedure to avoid interference from the secondary user to the primary one. The slow and fast fading scenarios can model situations in which secondary nodes and/or the primary user are stationary or moving terminals.}
	The numerical performance of each algorithm is computed using $10^4$ Monte Carlo runs. The data is generated following the model in (\ref{eq:energy_test}).
	The channel variance $\sigma^2_{n}$, which is equal to the quotient between the received power (without considering the noise) and the transmitted power, is modeled using the path-loss/log-normal shadowing model \cite{goldsmith2005wireless},
	\begin{equation}
		\sigma^2_{n}({\rm dB}) = K - 10\alpha \log_{10}(d_n/d_0) - \eta_n,\label{eq:var_h}    
	\end{equation}
	where $d_n$ is the distance between the source position and the $n$-th node position, $K$ (in dB) is the path-loss attenuation at a certain distance $d_0$, $\alpha$ is the path-loss exponent, and $\eta_{n}$ is a zero-mean Gaussian random variable with variance $\sigma^2_\eta$ which models the shadowing effect.
	In both slow and fast fading scenarios, the large-scale shadowing effect is assumed to be the same for all the sensing intervals, i.e., the variance (\ref{eq:var_h}) remains constant in the whole sensed interval.  
	Then, for each Monte Carlo run, $\sigma^2_n$ is sampled following (\ref{eq:var_h}). Thus, we obtain the average performance of the algorithms with respect to network channel variance distribution (induced by the source-node distance distribution, to be defined next, and the path-loss/shadowing model).
	
	The channel gains are determined as follows. In the slow fading case, it is assumed that the source and the sensors are static, so the channel gains $\{h_{n}\}_{n=1}^N$ (which are the same for all time-windows $l\in[1:L]$) are i.i.d. sampled from the PDF $\CN(0,\sigma^2_n)$ for each Monte Carlo run.
	On the other hand, in the fast fading scenario, it is assumed that the source and the sensors move, experiencing different small-scale fading realizations. This is modeled by sampling the channel gains $\{h_{n,l}\}_{n=1,l=1}^{N,L}$  i.i.d. (in both sensor index $n$ and time-window index $l$) from the PDF $\CN(0,\sigma^2_n)$, $n\in[1:N]$, for each Monte Carlo run. In both slow and fast fading, the channel amplitudes are Rayleigh distributed, a model representative of non-line-of-sight propagation scenarios.
	
	The chosen parameters for the simulation setup are shown in Table \ref{tab:param}. The selected propagation model parameters ($K$, $\alpha$, $d_0$ and $\sigma_\eta$) are typical for outdoors scenarios \cite[Ch. 2]{goldsmith2005wireless}.
	The signal-to-noise ratio is defined by $\text{SNR}\equiv \frac{\sqrt{M}P_s\bar{\sigma}^2}{W N_0}$, where $\bar{\sigma}^2$ is the average variance of the channels (\ref{eq:var_h}) across the nodes. Notice that the source power $P_s = \mbox{Tr}(\boldsymbol{\Sigma_s})/T$ will be varied to be consistent with the corresponding SNR.  The covariance matrix $\boldsymbol{\Sigma_s}$ is set to be a Toeplitz matrix with first row $\tfrac{P_s T}{M}[1,\rho,\dots,\rho^{M-1}]$, where $\rho= 0.5$.
	
\begin{figure}[htb!]
		\centering
		\if\mycmd0    
		\includegraphics[width=0.7\linewidth]{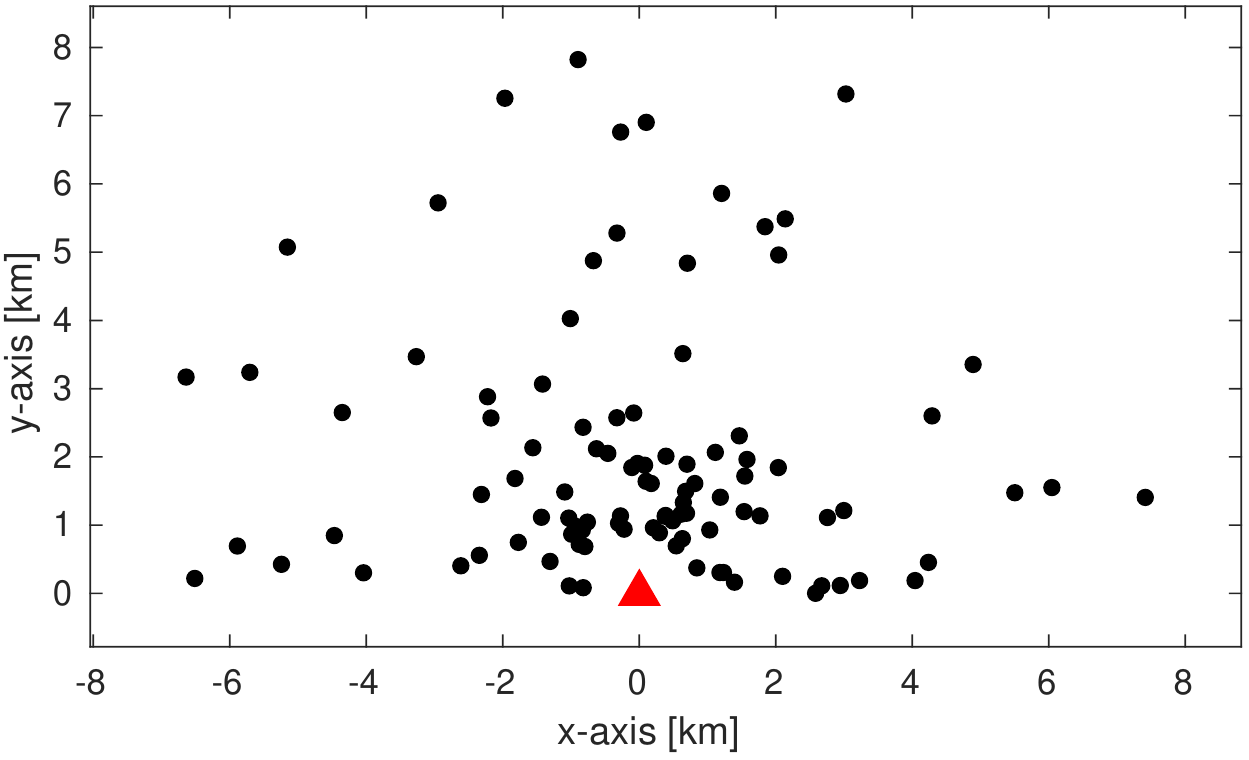}
		\else
		\includegraphics[width=0.5\linewidth]{20221013T110729NetGraphN100.pdf}
		\fi
		\caption{One realization of the sensor network for $N=100$ sensors represented with black dots. The source is depicted with a red triangle.}
		\label{fig:network_graphs}
	\end{figure}

	The source is assumed to be at $(0,0)$ m, and the location of each node in the plane $(x,y)$ is determined by $x =d_n\cos(\zeta_n)$ and $y =d_n\sin(\zeta_n)$.
	The distance between the source and the $n$-th node $d_n$ is assumed to be independently uniformly distributed in the log-scale\footnote{{The nodes distribution is selected such that nodes sensing relatively high power in a given frequency band wish to cooperate for detecting the source. This fits well in spectrum sensing schemes where cooperative nodes get some revenue for lending their sensing capabilities, as in \cite{kotobi2018secure}.}}, i.e., $\log_{10}(d_n/d_0) \sim \mathrm{U}[ \log_{10}(800\text{m}/d_0), \log_{10}(8000\text{m}/d_0)]$, and $\zeta_n$ is i.i.d. randomly selected from the uniform distribution $\mathrm{U}[0,\pi]$.
	In Fig. \ref{fig:network_graphs}, we show a realization of the sensor network.
	% Given the positions of the nodes, the network graph is built as follows: first, each node connects with its 3 nearest neighbors, and then, the edges between nodes are made symmetrical, meaning that if $\{i,j\}\in \mathcal{E}$, then $\{j,i\}$ is added to $\mathcal{E}$ if it was not already there. Finally, we check that the graph is connected.
	\begin{table}[bt!]
		\caption{Selected parameters for the simulation setup.}
		\centering
		\resizebox{\columnwidth}{!}{\begin{tabular}{cccccccccccccc}
				$K$(dB) &  $\alpha$ & $d_0$(m) & $\sigma_\eta$ & $W$(MHz) & $T$($\mu$s) & $M$ & $N_0$(dBm/Hz) & $N$ & $L$ \\
				\hline\hline
				-37 &  4 & 10 & 2 & 5 & 2 & 10 & -174 & 100  & 20 \\
		\end{tabular}}
		\label{tab:param}
	\end{table}
	The computation of the MLE in (\ref{eq:GLRT-SF}) and (\ref{eq:GLRT-FF}), needed for implementing both GLRT-SF and GLRT-FF algorithms, cannot be done through a closed form formula. So, we need a numerical procedure for solving this one-dimensional nonlinear problem with a bound constraint (the positiveness of the parameters $c_n$ and $d_n$) at each node. We use here a trust-region algorithm \cite{branch1999subspace} for which, depending on the scenario (fundamentally on the $\text{SNR})$ and the stopping criteria parameters (e.g. gradient tolerance and step tolerance), it typically takes 5-10 iterations to converge to a local maximum.
	
	We compare the performance of the proposed algorithms against the ones shown in Table II (see the definitions of the acronyms in the third column). The CSI-SF and CSI-FF detectors are the version of GLRT-SF and GLRT-FF, respectively, for which the parameters of the channel (channel state information, CSI) and the source are perfectly known. These statistics, of course, are unrealizable in practice and they are included only as a reference for the comparison. Note that some performance loss of the GLRT-based algorithms is expected with respect to these \emph{genie-aided} test statistics given the estimation errors of the unknown parameters.
	
	{In Table II, we also include the computational complexity of the test statistics considered in our simulations. It is straightforward to see that MD , SD, SC and SSC have complexity $\mathcal{O}(NL)$. % in terms of the energy measurements.
		Both ME and SSC require the computation of the eigenvalues of the sample covariance matrix of energy measurements at the nodes. This is computational demanding, requiring $\mathcal{O}(N^3)$ computations \cite{Pan_Chen_1999}. The cost of constructing the sample covariance matrix is $\mathcal{O}(N^2 L)$. Therefore, the total computational complexity of these algorithms is $\mathcal{O}(N^2(N+L))$.
		%\footnote{It is not included the cost of constructing the sample covariance matrix, which is $\mathcal{O}(N^2 L)$, under the assumption that $N\geq L$.}.  
		The complexity for the genie-aided cases CSI-SF and CSI-FF in which the parameters $c_n$ and $d_n$ are known is $\mathcal{O}(NL)$. When those parameters are unknown, GLRT-SF and GLRT-FF estimate them. Then, we need to add the cost of estimating the unknown parameters $c_n$ and $d_n$ by a maximization procedure. It is important to note that the problem of obtaining the $N$ optimal values of $c_n$ or $d_n$ can be decomposed into $N$ scalar optimization problems, which are obviously significantly less complex than a single $N-$dimensional optimization problem. This is a consequence of the factorized PDF obtained by our approximations.  In our case, we used a trust-region algorithm which is based on the conjugate gradient method. This is an iterative method whose complexity depends on the accuracy required for the solution. It is known that if a tolerance of $\xi$ is the solution required, the worst-case complexity is $\mathcal{O}(\xi^{-\upsilon})$ where $\upsilon\in[1,2]$ depending on the properties of the function to be optimized \cite{Gratton_Royer_Vicente_Zhang_2018}.  In practice, the average complexity seems to be significantly lower.  In our experience, for the problem considered in this paper and as explained above, the convergence to a local maximum was very fast, and the complexity of both  GLRT-SF and GLRT-FF is dominated by $\mathcal{O}(NL)$.}
	
	%The Newton-Rapshon procedure is the method of choice whose complexity depends on the accuracy required because of its iterative nature. For an $n-$ digit precision, the usual time complexity of Newton-Raphson is $\mathcal{O}(A(n)\log n )$, where $A(n)$ is a fixed cost that depends on the function to be optimized. For example, in the case of the GLRT-SF this will depend on the functional form of (\ref{eq:aprox_SF}). If the required precision is not large enough, the cost will be dominated by the term $\mathcal{O}(NL)$.}

The test statistic MD is the mean detector, i.e., the average of all network measurements, and is equivalent to the equal gain combining (EGC), typically used in low SNR regimes \cite{niu2006fusion}. The test statistic SD is the square detector. We also include two eigenvalue-based detectors (ME and SSE) which are typically used in the present scenario. These detectors naturally consider the statistical dependence of the observations under $\Hip_1$ at different sensor nodes, introduced by the random signal source, given that they are based on the eigenvalues $\{\lambda_n\}_{n=1}^N$ of the sample covariance matrix of the observations $\{\mathbf{E}_l\}_{l=1}^L$. Nevertheless, judging by the analysis in the previous section, and the following numerical results, the statistical dependence appears to have a negligible impact in the detectors' performance when considering energy observations.
%We also remark that eigenvalue-based detectors are not well suited for fully-distributed scenarios with scarce resources, due to the distributed implementation of these algorithms require lots of communication and energy resources \cite{maya2021fully,penna_decentralized_2015}.
Finally, we also include the SC detector \cite{digham2003energy}, which selects the highest average measured energy among all the sensors, and the SSC detector \cite{maya2022fading}.

\begin{table}[bt!]
	\caption{Test statistics to be compared with GLRT-SF and GLRT-FF. $\dagger$ are genie-aided detectors.}
	\resizebox{\columnwidth}{!}{\begin{tabular}{|c|p{.3\linewidth}|p{.3\linewidth}|c|}
			\hline
			Label    &  Test statistic/Ref. &  Detector name / Observation &  Complexity\\
			\hline
			GLRT-SF    & Eq. (\ref{eq:GLRT-SF}) & GLRT-based detector for slow fading (SF) & $\mathcal{O}(NL)$\\
			\hline
			GLRT-FF    & Eq. (\ref{eq:GLRT-FF}) & GLRT-based detector for fast fading (FF) & $\mathcal{O}(NL)$\\
			\hline
			CSI-SF$\dagger$    & Eq. (\ref{eq:GLRT-SF}) using the true value $c_n$ instead of $\hat{c}_n$, $n\in[1,N]$. & CSI detector for slow fading (SF). & $\mathcal{O}(NL)$\\
			\hline
			CSI-FF$\dagger$    & Eq. (\ref{eq:GLRT-FF}) using the true value $d_n$ instead of $\hat{d}_n$, $n\in[1,N]$. & CSI detector for fast fading (FF). & $\mathcal{O}(NL)$\\
			\hline
			MD    & $1/(NL) \sum_{n=1,l=1}^{N,L} {E}_{n,l}$&  Mean detector. &$\mathcal{O}(NL)$\\
			\hline
			SD    & $1/(NL) \sum_{n=1,l=1}^{N,L} {E}_{n,l}^2$ &  Square detector. &$\mathcal{O}(NL)$\\
			\hline
			ME & $\lambda_\text{max}$, \cite{taherpour_multiple_2010}    & Maximum eigenvalue detector. $\lambda_\text{max}$ is the maximum eigenvalue of the sample covariance matrix. & $\mathcal{O}(N^2(N\!+\!L))$\\
			\hline
			SSE & $\sum_{n=1}^N -\log\lambda^+_n + \lambda^+_n$, \cite{zhang_multi-antenna_2010}    & Subspace eigenvalue detector. $\lambda^+_n \equiv \max(0,\lambda_n-1)$ & $\mathcal{O}(N^2(N\!+\!L))$\\
			\hline
			SC & $\max_{n\in[1,N]} m_{z,n}$,\cite{digham2003energy}    & Selection combining detector & $\mathcal{O}(NL)$\\
			\hline
			SSC & $\sum_{n=1}^N \max(m_{z,n},0)^2$,\cite{maya2022fading}    & Selection square combining detector &$\mathcal{O}(NL)$\\
			\hline
	\end{tabular}}
	\label{tab:stats}
\end{table}

\begin{figure}[]
	% Figures generated using the script (in Curie): %/home/users/jumaya/Simulations/SourceSTD_v1.m
	%/home/users/jumaya/Simulations/SourceSTD_v1_plots.m
	% Path to folder: /home/users/jumaya/Simulations/Figs_STD_v1/
	\centering
	\if\mycmd0    
	\includegraphics[width=\linewidth]{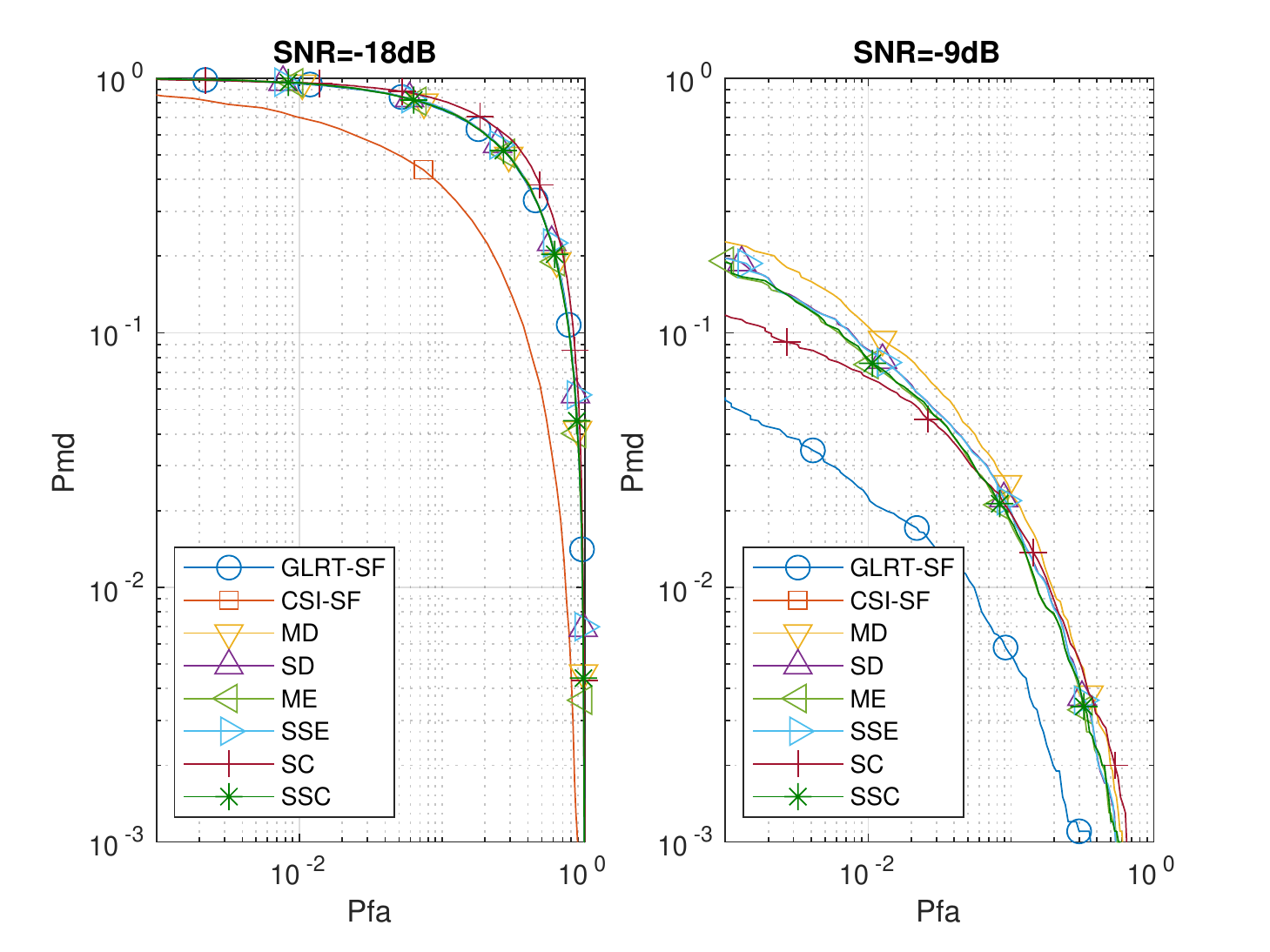}
	\includegraphics[width=\linewidth]{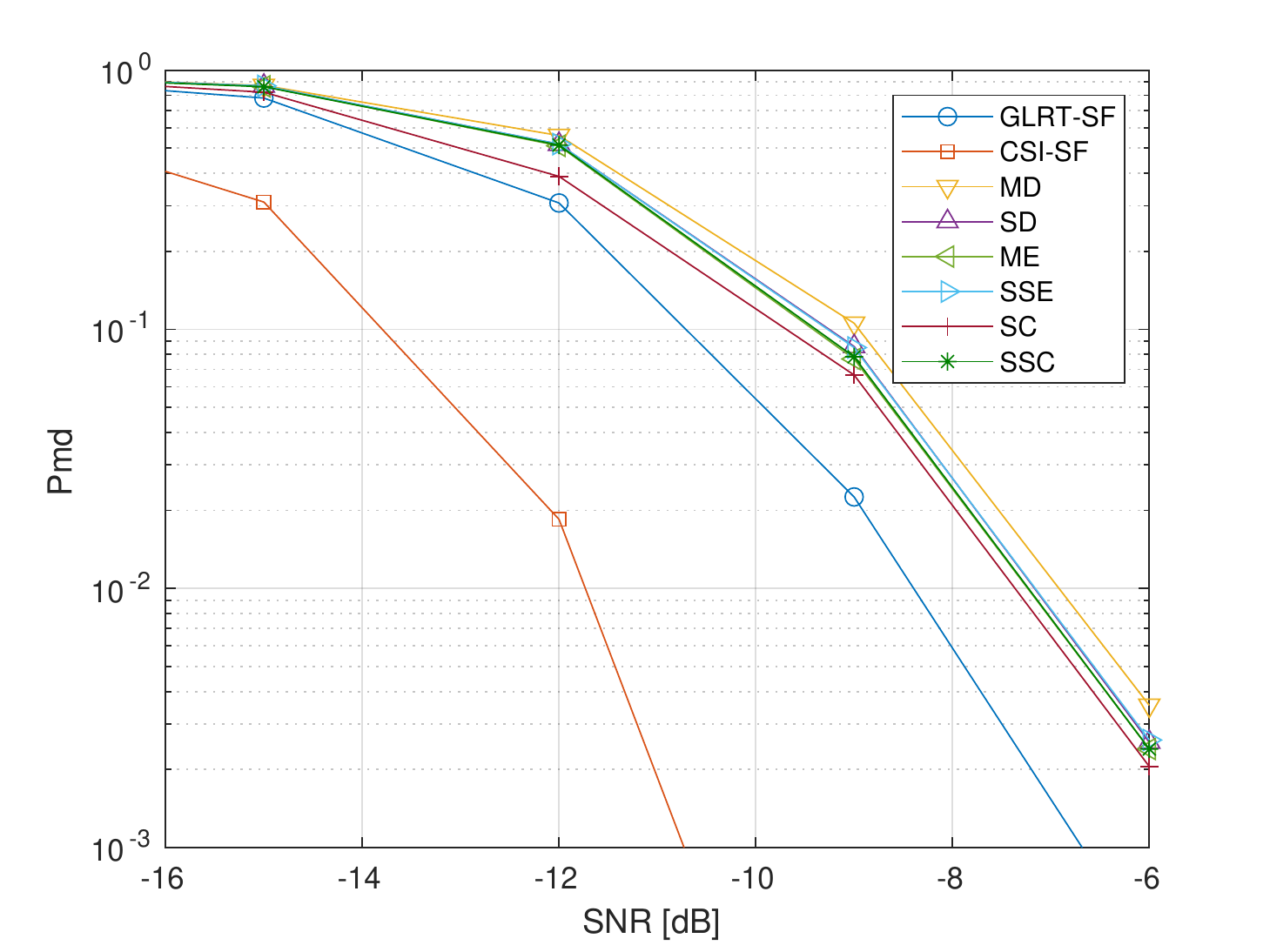}
	\else
	\includegraphics[width=.48\linewidth]{20220725T094343_CROC_sf}
	\includegraphics[width=.48\linewidth]{20220725T094343_Pmd_SNR_L20_WT10_sf}
	\fi
	\caption{Performance of the algorithms for slow fading channels. Top: Complementary receiver operating characteristics for SNR $=-18$dB (left) and SNR $=-9$dB (right). Notice that on the right plot, the CSI-SF curve is not seen because it falls outside below the range shown. Bottom: Miss detection probabilities vs SNR for false alarm probability $P_\text{fa}=0.01$.}
	\label{fig:Performance_sf}
\end{figure}

\begin{figure}[]
	% Figures generated using the script (in Curie): %/home/users/jumaya/Simulations/SourceSTD_v1.m
	%/home/users/jumaya/Simulations/SourceSTD_v1_plots.m
	% Path to folder: /home/users/jumaya/Simulations/Figs_STD_v1/    \centering
	\centering
	\if\mycmd0    
	\includegraphics[width=\linewidth]{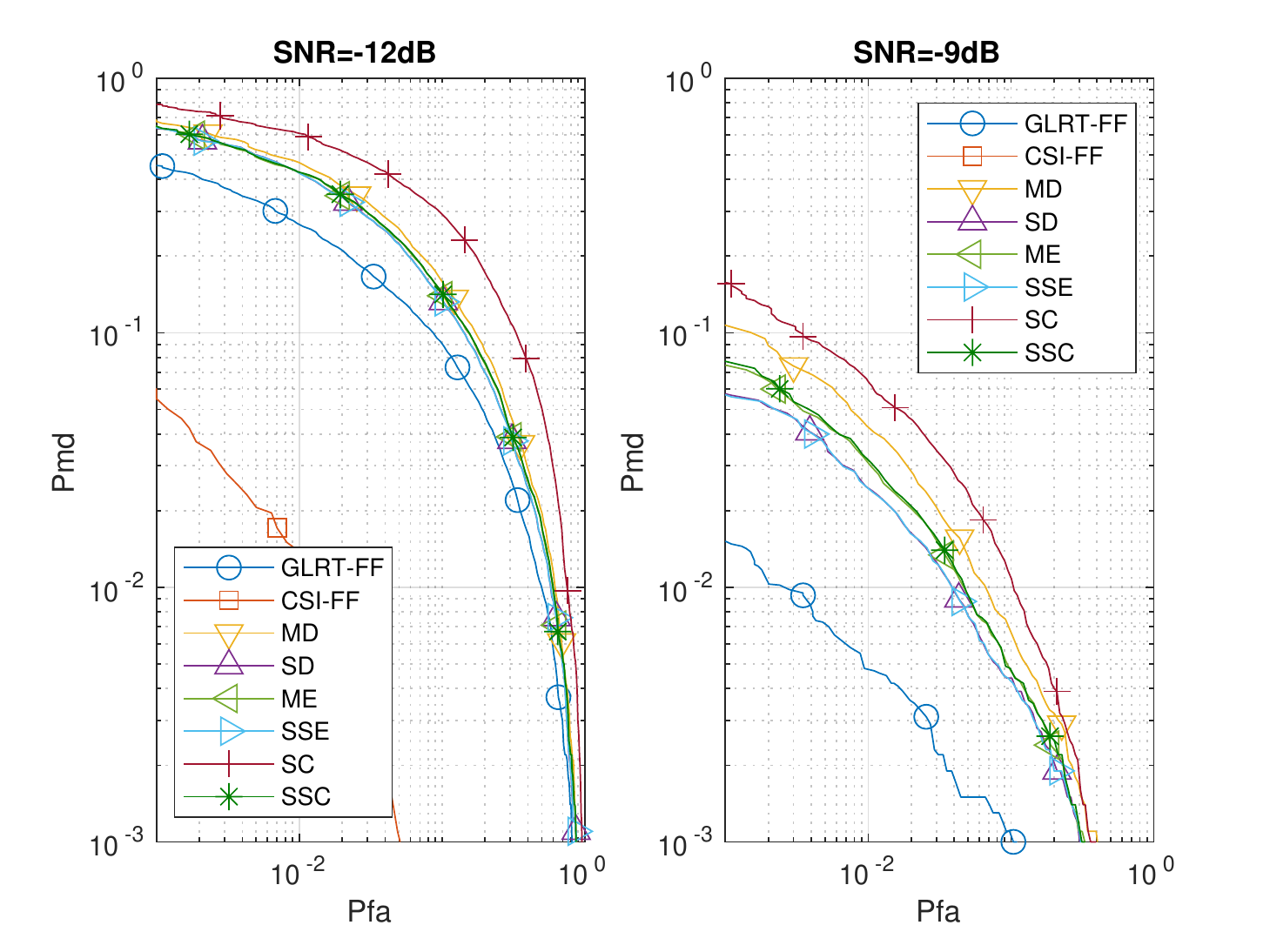}
	\includegraphics[width=\linewidth]{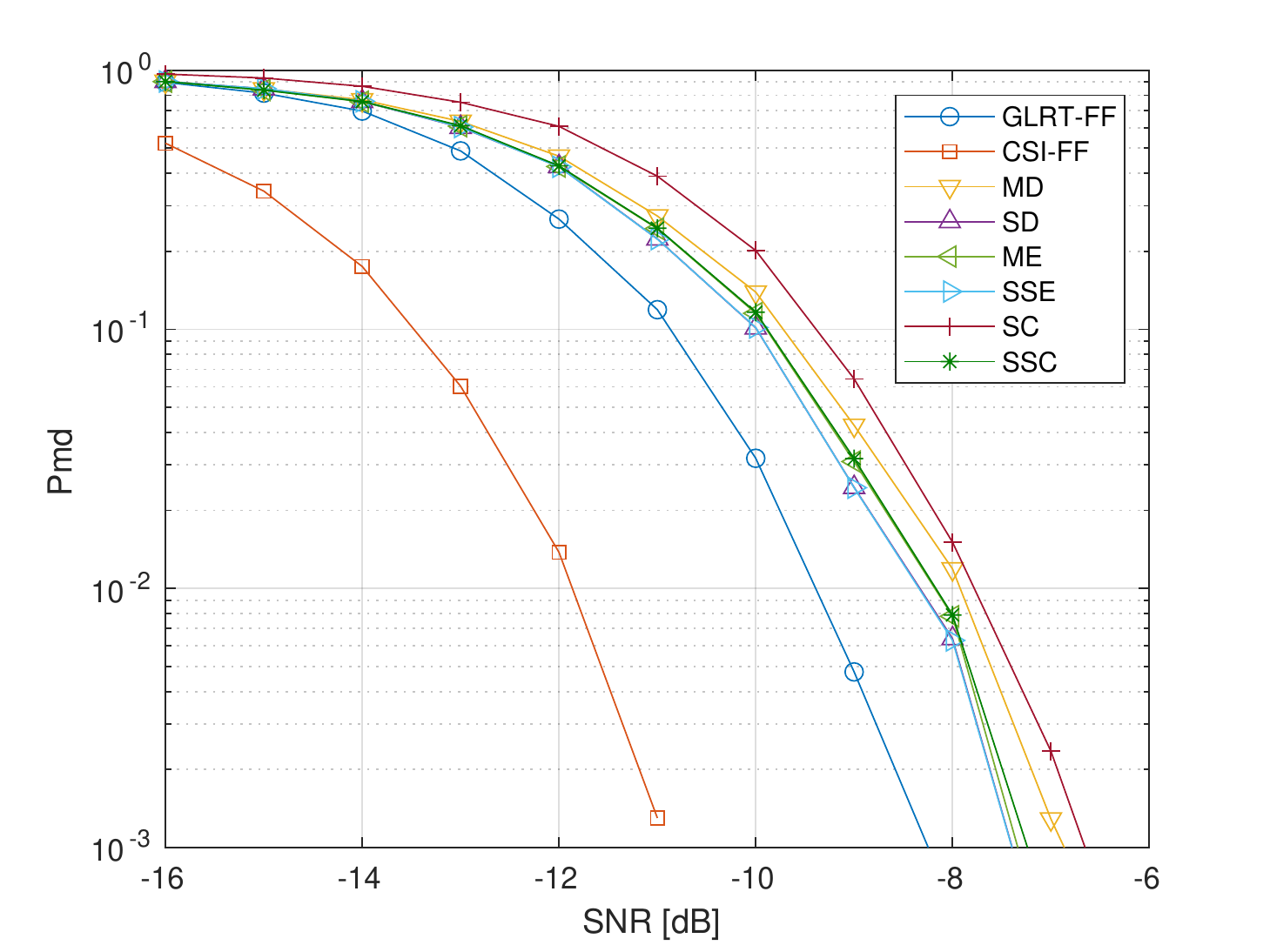}
	\else
	\includegraphics[width=.48\linewidth]{20220725T192931_CROC_ff}
	\includegraphics[width=.48\linewidth]{20220725T192931_Pmd_SNR_L20_WT10_ff}
	\fi
	\caption{Performance of the algorithms for fast fading channels. Top: Complementary receiver operating characteristics for SNR $=-12$dB (left) and SNR $=-9$dB (right). Notice that on the right plot, the CSI-FF curve is not seen because it falls outside below the range shown. Bottom: Miss detection probabilities vs SNR for false alarm probability $P_\text{fa}=0.01$.}
	\label{fig:Performance_ff}
\end{figure}

In Fig. \ref{fig:Performance_sf} and \ref{fig:Performance_ff} (top figures), we plot the complementary receiver operating characteristic (CROC), i.e., the miss detection probability against the false alarm probability (both defined in Section II), for the algorithms in Table II, in the slow fading and the fast fading scenario, respectively, for the indicated SNRs. Additionally, in the respective bottom figures, we plot the miss detection probability against the SNR, when $P_\text{fa}=10^{-2}$. As expected, the genie-aided test statistics outperform the rest of the algorithms. Among the test statistics capable of being implemented in practice, GLRT-SF and GLRT-FF, achieve the best performance for each scenario. These gains seem to increase for large values of SNRs. This is important, because among the other methods there are some that explicitly exploit information about spatial correlation (as ME and SSE) and others that do not (as MD and SC). In addition, the proposed detectors demand similar communication resources to the simplest ones, and provide better performance than the ones that make use of the spatial correlation information among the nodes and also demand more network communication resources \cite{Wiesel_Hero_2012}.

\begin{figure}[]
	% Figures generated using the script (in Curie): %/home/users/jumaya/Simulations/SourceSTD_v1.m
	%/home/users/jumaya/Simulations/SourceSTD_v1_plots.m
	% Path to folder: /home/users/jumaya/Simulations/Figs_STD_v1/    \centering
	\centering
	\if\mycmd0    
	\includegraphics[width=\linewidth]{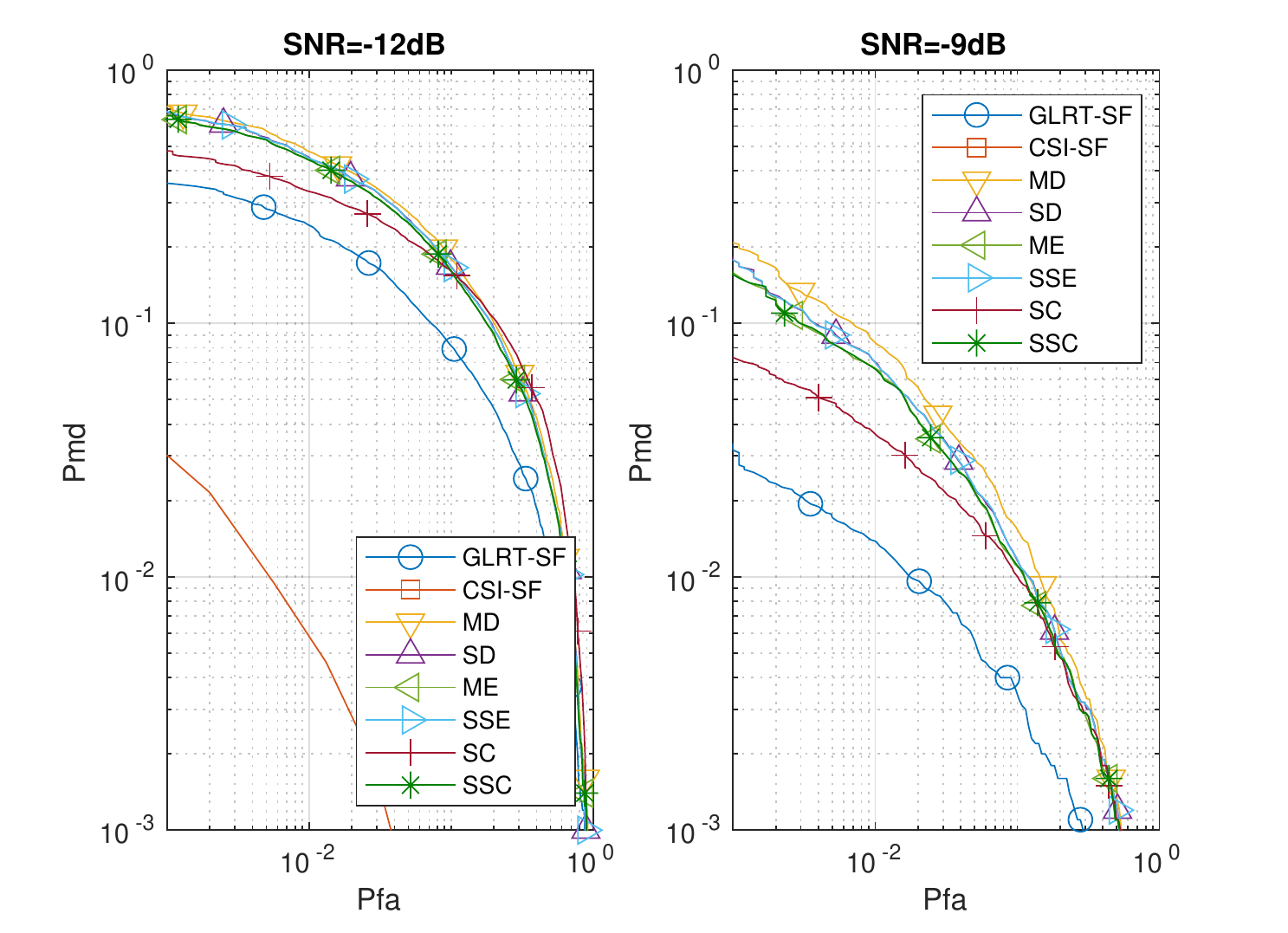}
	\includegraphics[width=\linewidth]{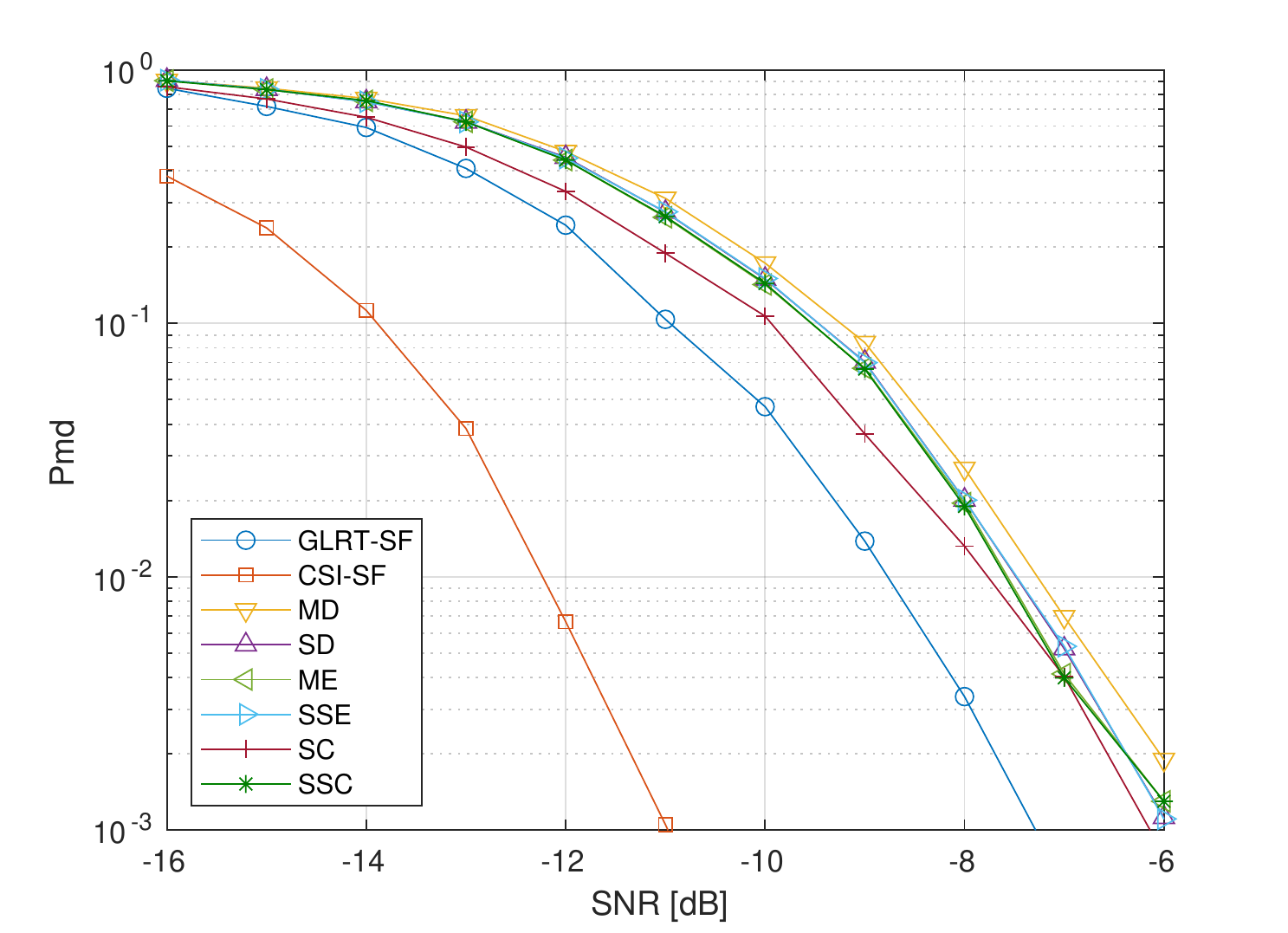}
	\else
	\includegraphics[width=.48\linewidth]{20230629T172330_CROC_sf}
	\includegraphics[width=.48\linewidth]{20230629T172330_Pmd_SNR_L20_WT15_sf}
	\fi
	\caption{OFDM signal: Performance of the algorithms for slow fading channels when the source follows an OFDM modulation. Top: Complementary receiver operating characteristics for SNR $=-12$dB (left) and SNR $=-9$dB (right). Notice that on the right plot, the CSI-SF curve is not seen because it falls outside below the range shown. Bottom: Miss detection probabilities vs SNR for false alarm probability $P_\text{fa}=0.01$.}
	\label{fig:Performance_ofdm}
\end{figure}

Finally, in Fig. \ref{fig:Performance_ofdm}, we also consider a case where the source signal is not Gaussian but a OFDM signal.  {We consider a NB-IoT and LTE-M scenarios, where typical small FFT sizes are used, such as 12 or 24, to allow for narrower subcarrier spacing and more efficient transmission of low-rate IoT data \cite{Wang_Lin_Adhikary_Grovlen_Sui_Blankenship_Bergman_Razaghi_2017}. Notice that also a setting with such a small OFDM symbol length $M$ is interesting as can be thought as a limiting case to our assumption that $M\approx WT$ is large. In particular, in our case,  the number of subcarriers is set to 12 (a resource block in Long Term Evolution cellular systems) and the cyclic prefix is defined to be 3 samples. Then, the OFDM symbol has a length of $M=15$. We consider a 64-QAM as the subscarriers modulation. We can see that similar conclusions as for the Gaussian signal can be drawn, which validates the theoretical approach used in this work for this particular setup.}
\section{Concluding remarks}

We have considered the problem of energy-based distributed detection of a stochastic Gaussian radio source signal, for which the measurements at each sensor node present spatial and temporal correlation. Using the closed-form expression for the CF of the joint PDF of the energy measurement at each sensor site, we computed an approximation of the joint PDF (when the source signal is present), under the assumption of a large time-bandwidth product, and for both wireless scenarios of common interest: slow and fast fading. We also provided two deviation bounds for the obtained approximation respect to the true intractable joint PDF. These bounds prove to be exponential on the number of nodes in the network. The bounds indicate that the spatial and temporal correlation of the measurements at each node site are not extremely critical given that the exact joint PDF can be tightly approximated by a factorized PDF. These PDF approximations were used to implement likelihood ratio tests that show performance gains with respect to other usual schemes typically used in practice. Moreover, the product nature of the obtained PDF was shown to be valuable in implementing simpler cooperative schemes in distributed scenarios. {It also can be an asset for designing simple quantization schemes relying on the local PDF at each node in order to cope, among other things, with the usual reporting channel impairments between the nodes and/or the FC. Another possible line of work could include the use of the approximation to analyze the outage probability of different receiving techniques of large MIMO systems with correlation on the transmitter and receiver antennas (see for example the results in \cite{Morales-Jimenez_Paris_Entrambasaguas_Wong_2011} and \cite{Ermolova_Tirkkonen_2012}). The application of the obtained results for characterizing performance of distributed radar systems \cite{Yang_Lai_Jakobsson_Yi_2023}, can also be explored in the future.}

\appendices
\section{Proof of Lemma \ref{lemma:CF}}
\label{ap:CF}

We can write (under both hypotheses $\mathcal{H}_0$ and $\mathcal{H}_1$):
\begin{equation}
	\Psi(\boldsymbol\omega|\mathbf{h})=\mathbb{E}\left[e^{j\frac{1}{\beta(M)}\sum_{n=1}^N\omega_n\|\mathbf{y}_{n}\|^2}\right].
\end{equation}
Assuming that $\boldsymbol\Sigma_{\mathbf{s}}=\mathbf{Q}\boldsymbol\Lambda\mathbf{Q}^H$ where $\mathbf{Q}$ is unitary and $\boldsymbol\Lambda=\mbox{diag}(\lambda_{1},\lambda_{2},\dots,\lambda_{M})$ with $\lambda_{m}$ the $m$-largest positive eigenvalue of $\boldsymbol\Sigma_{\mathbf{s}}$, we can define $\tilde{\mathbf{y}}\equiv \left(\mathbf{I}_N\otimes\mathbf{Q}\right)\mathbf{y}$. It is immediate to see that under $\mathcal{H}_0$, $\tilde{\mathbf{y}}\sim\mathcal{C}\mathcal{N}(\mathbf{0},\sigma_v^2\mathbf{I}_{NM})$ and under $\mathcal{H}_1$, $\tilde{\mathbf{y}}\sim\mathcal{C}\mathcal{N}(\mathbf{0},\mathbf{h}\mathbf{h}^H\otimes\boldsymbol\Lambda+\sigma_v^2\mathbf{I}_{NM})$. Moreover, we have that
\begin{equation}
	\Psi(\boldsymbol\omega|\mathbf{h})=\mathbb{E}\left[e^{j\frac{1}{\beta(M)}\sum_{n=1}^N\omega_n\|\tilde{\mathbf{y}}_{n}\|^2}\right]=\mathbb{E}\left[e^{j\frac{1}{\beta(M)}\tilde{\mathbf{y}}^H\mathbf{A}_{\boldsymbol\omega}\tilde{\mathbf{y}}}\right],
\end{equation}
where we have used that $\mathbf{Q}$ preserves the norm and with  $\mathbf{A}_{\boldsymbol\omega}=\mbox{diag}(\omega_1,\omega_2,\dots,\omega_N)\otimes \mathbf{I}_M$. As under both hypotheses, $\tilde{\mathbf{y}}$ is a complex and circular Gaussian random vector, it is easy to obtain that under $\mathcal{H}_0$:
\begin{equation}
	\Psi^0(\boldsymbol{\omega}|\mathbf{h})=\prod_{n=1}^N \frac{1}{\left(1-j\frac{\omega_n\sigma_v^2}{\beta(M)}\right)^M}.
\end{equation}
Under $\mathcal{H}_1$ we obtain:
\begin{equation}
	\Psi^1(\boldsymbol{\omega}|\mathbf{h})=\frac{1}{\Big|\mathbf{I}_{NM}-j\frac{\left(\mathbf{h}\mathbf{h}^H\otimes\boldsymbol\Lambda+\sigma_v^2\mathbf{I}_{NM}\right)\mathbf{A}_{\boldsymbol{\omega}}}{\beta(M)}\Big|}
\end{equation}
Note that we can write:
\begin{equation}
	\left(\mathbf{h}\mathbf{h}^H\otimes\boldsymbol\Lambda\right)\mathbf{A}_{\boldsymbol{\omega}}
	=\left(\mathbf{h}\mathbf{h}^H\mbox{diag}(\omega_1,\omega_2,\dots,\omega_N)\right)\otimes\boldsymbol\Lambda,
	\label{eq:kronecker_comm}
\end{equation}
where we have used that $(\mathbf{A}\otimes\mathbf{B})(\mathbf{C}\otimes\mathbf{D})=\left(\mathbf{A}\mathbf{C}\right)\otimes\left(\mathbf{B}\mathbf{D}\right)$. Consider now the \emph{commutation matrix} $\mathbf{K}_{N,M}$ \cite{Magnus_Neudecker_1979}. This matrix is a permutation matrix of size $NM\times NM$ with several interesting properties. The one we need is the following: for square matrices $\mathbf{A}\in\mathbb{C}^{N\times N}$, $\mathbf{B}\in\mathbb{C}^{M\times M}$, we have $\mathbf{A}\otimes\mathbf{B}=\mathbf{K}_{N,M}\left(\mathbf{B}\otimes\mathbf{A}\right)\mathbf{K}_{N,M}^T$. Using this matrix and the result of (\ref{eq:kronecker_comm}) we can express $\mathbf{I}_{NM}-j\frac{\left(\mathbf{h}\mathbf{h}^H\otimes\boldsymbol\Lambda+\sigma_v^2\mathbf{I}_{NM}\right)\mathbf{A}_{\boldsymbol{\omega}}}{\beta(M)}$ as:
\if\mycmd1
\begin{equation}
	\!\!\!\!\!\mathbf{K}_{N,M}\!\left(\!\mathbf{I}_{NM}\!-\!j\frac{\boldsymbol\Lambda\otimes\left(\mathbf{h}\mathbf{h}^H\boldsymbol{\Omega}\right)\!+\!\sigma_v^2\mathbf{K}_{N,M}^T\left(\boldsymbol{\Omega}\otimes\mathbf{I}_M\right)\mathbf{K}_{N,M}}{\beta(M)}\!\right)\mathbf{K}_{N,M}^T,
	\label{eq:CF_after_comm}
\end{equation}
\else
\begin{multline}
	\!\!\!\!\!\mathbf{K}_{N,M}\!\left(\!\mathbf{I}_{NM}\!-\!j\frac{\boldsymbol\Lambda\otimes\left(\mathbf{h}\mathbf{h}^H\boldsymbol{\Omega}\right)\!+\!\sigma_v^2\mathbf{K}_{N,M}^T\left(\boldsymbol{\Omega}\otimes\mathbf{I}_M\right)\mathbf{K}_{N,M}}{\beta(M)}\!\right)\nonumber\\
	\times\mathbf{K}_{N,M}^T,
	\label{eq:CF_after_comm}
\end{multline}
\fi
where we have defined $\boldsymbol{\Omega}=\mbox{diag}(\omega_1,\omega_2,\dots,\omega_N)$. In first place note that, as $\mathbf{K}_{N,M}^T=\mathbf{K}_{M,N}$, $\mathbf{K}_{N,M}^T\left(\boldsymbol{\Omega}\otimes\mathbf{I}_M\right)\mathbf{K}_{N,M}$ transforms into $\mathbf{I}_M\otimes\boldsymbol{\Omega}$. In second place, the matrix $\boldsymbol\Lambda\otimes\left(\mathbf{h}\mathbf{h}^H\boldsymbol{\Omega}\right)$ is a block-diagonal matrix with diagonal blocks given by $\lambda_{m}\mathbf{h}\mathbf{h}^H\boldsymbol{\Omega}$ with $m=1,\dots,M$. As a consequence, and using the fact that $\Big|\mathbf{K}_{N,M}\Big|\Big|\mathbf{K}_{N,M}^T\Big|=1$, (\ref{eq:kronecker_comm}) can be put as:
\begin{equation}
	\Psi^1(\boldsymbol{\omega}|\mathbf{h})=\frac{1}{\prod_{m=1}^M\Big|\mathbf{I}_N-j\frac{\sigma_v^2\boldsymbol{\Omega}}{\beta(M)}-j\frac{\lambda_{m}\mathbf{h}\mathbf{h}^H\boldsymbol{\Omega}}{\beta(M)}\Big|}.
	\label{eq:CF_2}
\end{equation}
Finally, using the fact that $|\mathbf{A}+\mathbf{u}\mathbf{v}^H|=|\mathbf{A}|(1+\mathbf{v}^H\mathbf{A}^{-1}\mathbf{u})$, for every invertible matrix $\mathbf{A}$ and vectors $\mathbf{u},\mathbf{v}$ of appropriate size, we get:
\begin{equation}
	\Psi^1(\boldsymbol{\omega}|\mathbf{h})=\frac{\prod_{m=1}^M\Big|\mathbf{I}_N-j\frac{\sigma_v^2\boldsymbol{\Omega}}{\beta(M)}\Big|^{-1}}{\prod_{m=1}^M\Big(1-j\frac{\lambda_{m}}{\beta(M)}\mathbf{h}^H\left(\mathbf{I}_N-j\frac{\sigma_v^2\boldsymbol{\Omega}}{\beta(M)}\right)^{-1}\mathbf{h}\Big)}.
	\label{eq:CF_3}
\end{equation}
As $\boldsymbol{\Omega}$  is a diagonal matrix we can easily get the result of the Lemma.

\section{Proof of Lemma \ref{lemma:fading}}
\label{ap:fading}
As $\hat{p}^{1}(\mathbf{E}|\mathbf{h})$ in (\ref{eq:approx_pdf1}) is product PDF  and the channel gains are independent we can write:
\begin{equation}
	\label{eq:first_exp}
	\mathbb{E}\left[\hat{p}^{1}(\mathbf{E}|\mathbf{h})\right]=\prod_{n=1}^{N}\int_{0}^{\infty} \hat{p}^{1}(E_n||h_n|^2)p_n(|h_n|^2)d|h_n|^2,
\end{equation}
where
\begin{multline}
	\label{eq:approx_pdf_marg}
	\!\!\hat{p}^1(E_n||h_n|^2)= \tfrac{\beta(M)}{\sigma_v^2}\exp\left[-\frac{\beta(M)}{\sigma_v^2}\Big(E_{n}+\frac{\mbox{Tr}\left({\boldsymbol{\Sigma}_{\mathbf{s}}}\right)|h_n|^2}{\beta(M)}\Big)\right]\\
	\times\left(\tfrac{\beta(M)E_{n}}{\mbox{Tr}\left({\boldsymbol{\Sigma}_{\mathbf{s}}}\right)|h_n|^2}\right)^{\frac{M-1}{2}} I_{M-1}\left(\tfrac{2}{\sigma_v^2}\sqrt{\beta(M)\mbox{Tr}\left({\boldsymbol{\Sigma}_{\mathbf{s}}}\right)|h_n|^2E_{n}}\right)\nonumber,
\end{multline}
for $n\in[1:N]$. Then, we need only to consider the computation of:
\begin{equation}
	\mathbb{E}\left[\hat{p}^{1}(E_n||h_n|^2)\right]=\int_{0}^{\infty} \hat{p}^1(E_n||h_n|^2) p_n(|h_n|^2)d|h_n|^2
	\label{eq:second_exp}
\end{equation}
Let us define the CF $\tilde{\psi}^{1}_n(\omega_n)$  of $\mathbb{E}\left[\hat{p}^{1}(E_n||h_n|^2)\right]$ with $n\in[1:N]$. Using Fubini's Theorem \cite{Billingsley_1995} we can write:
\if\mycmd1
\begin{eqnarray}
	\label{eq:CF_aux_fad}
	\tilde{\psi}^{1}_n(\omega_n)\!\!\!\!\!&=&\!\!\!\!\!\int_{0}^\infty \mathbb{E}\left[\hat{p}^{1}(E_n||h_n|^2)\right]e^{j\omega_n E_n}dE_n=\int_{0}^{\infty}\!\!\left[\int_{0}^{\infty}\!\!\hat{p}^1(E_n||h_n|^2)e^{j\omega_n E_n}dE_n\!\right]\! p_n(|h_n|^2)d|h_n|^2\nonumber\\
	&=&\!\!\!\!\!\ \mathbb{E}\left[\hat{\psi}^1(\omega_n|h_n)\right]
\end{eqnarray}
\else
\begin{eqnarray}
	\label{eq:CF_aux_fad}
	\tilde{\psi}^{1}_n(\omega_n)\!\!\!\!\!&=&\!\!\!\!\!\int_{0}^\infty \mathbb{E}\left[\hat{p}^{1}(E_n||h_n|^2)\right]e^{j\omega_n E_n}dE_n\nonumber\\
	&=&\!\!\!\!\!\int_{0}^{\infty}\!\!\left[\int_{0}^{\infty}\!\!\hat{p}^1(E_n||h_n|^2)e^{j\omega_n E_n}dE_n\!\right]\! p_n(|h_n|^2)d|h_n|^2\nonumber\\
	&=&\!\!\!\!\!\ \mathbb{E}\left[\hat{\psi}^1(\omega_n|h_n)\right]
\end{eqnarray}
\fi
where $\hat{\psi}^1(\omega_n|h_n)$ was defined in  (\ref{eq:approx_3}). It is easy to show that $\mathbb{E}\left[\hat{\psi}^1(\omega_n|h_n)\right]$ has a closed form when $p_n(|h_n|^2)$ is as in (\ref{eq:Rayleigh}):
\begin{equation}
	\tilde{\psi}^{1}_n(\omega_n)=\frac{1}{\left(1-j\frac{\sigma_v^2\omega_n}{\beta(M)}\right)^{M-1}}\frac{1}{\left(1-j\frac{\sigma_v^2+\mbox{Tr}(\boldsymbol{\Sigma_s})\sigma_n^2}{\beta(M)}\omega_n\right)}.
	\label{eq:CF_aux_fad2}
\end{equation}
Notice that  $\tilde{\psi}^{1}_n(\omega_n)$ is the product of two well-known CFs. In first place, the first term in the RHS of (\ref{eq:CF_aux_fad}) is the CF of a Gamma distribution with shape $M-1$ and scale $\frac{\sigma_v^2}{\beta(M)}$. Similarly, the second term in the RHS of (\ref{eq:CF_aux_fad}) is the CF of an exponential distribution with scale parameter $\frac{\sigma_v^2+\mbox{Tr}\left(\boldsymbol{\Sigma_s}\right)\sigma_n^2}{\beta(M)}$. Clearly the PDF $\mathbb{E}\left[\hat{p}^{1}(E_n||h_n|^2)\right]$ which corresponds to the CF $\tilde{\psi}^{1}_n(\omega_n)$ is the convolution of the above described pdfs. The calculation can be easily done to obtain for each $n\in[1:N]$:
\if\mycmd1
\begin{equation}
	\mathbb{E}\left[\hat{p}^{1}(E_n||h_n|^2)\right]=\tfrac{\beta(M)}{\Gamma(M-1)}\tfrac{\left(\sigma_v^2+\mbox{Tr}\left(\boldsymbol{\Sigma_s}\right)\sigma_n^2 \right)^{M-2}}{\left(\mbox{Tr}\left(\boldsymbol{\Sigma_s}\right)\sigma_n^2\right)^{M-1}}\exp\!\left(\!-\tfrac{\beta(M)E_n}{\sigma_v^2+\mbox{Tr}\left(\boldsymbol{\Sigma_s}\right)\sigma_n^2}\!\right)\!\gamma\left(\!M\!-\!1,\!\tfrac{\beta(M)\mbox{Tr}\left(\boldsymbol{\Sigma_s}\right)\sigma_n^2}{\sigma_v^2\left(\sigma_v^2+\mbox{Tr}\left(\boldsymbol{\Sigma_s}\right)\sigma_n^2\right)}E_n\!\right),
\end{equation}
\else
\begin{multline}
	\mathbb{E}\left[\hat{p}^{1}(E_n||h_n|^2)\right]=\tfrac{\beta(M)}{\Gamma(M-1)}\tfrac{\left(\sigma_v^2+\mbox{Tr}\left(\boldsymbol{\Sigma_s}\right)\sigma_n^2 \right)^{M-2}}{\left(\mbox{Tr}\left(\boldsymbol{\Sigma_s}\right)\sigma_n^2\right)^{M-1}}\times\nonumber\\
	\exp\!\left(\!-\tfrac{\beta(M)E_n}{\sigma_v^2+\mbox{Tr}\left(\boldsymbol{\Sigma_s}\right)\sigma_n^2}\!\right)\!\gamma\left(\!M\!-\!1,\!\tfrac{\beta(M)\mbox{Tr}\left(\boldsymbol{\Sigma_s}\right)\sigma_n^2}{\sigma_v^2\left(\sigma_v^2+\mbox{Tr}\left(\boldsymbol{\Sigma_s}\right)\sigma_n^2\right)}E_n\!\right),
\end{multline}
\fi
where $\gamma(\alpha,x)\equiv\int_{0}^{x}t^{\alpha-1}e^{-t}dt$, $\alpha,x\geq 0$ is the incomplete Gamma function. The final result can be easily obtained from here.

\section{Proof of Lemma \ref{lemma:bound_z}}
\label{ap:z}
As $\mathfrak{Re}(z_m(\boldsymbol{\omega}))\leq 0$ for all $\boldsymbol{\omega}\in\mathbb{R}^N$ and $m\in[1:M]$ we have:
\begin{equation}
	\label{eq:z_less_1}
	\left|\frac{1}{1-z_m(\boldsymbol{\omega})}\right|\leq 1,\  \left|e^{z_m(\boldsymbol{\omega})}\right|\leq 1, \forall \boldsymbol{\omega}\in\mathbb{R}^N,\ m\in[1:M].
\end{equation}
We can write:
\if\mycmd1
\begin{equation}
	\label{eq:product_bound_z}
	\!\!\!\!\!\!\prod_{m=1}^M\tfrac{1}{1-z_m(\boldsymbol{\omega})}-\prod_{m=1}^Me^{z_m(\boldsymbol{\omega})}=
	\left(\tfrac{1}{1-z_M(\boldsymbol{\omega})}-e^{z_M(\boldsymbol{\omega})}\right)\prod_{m=1}^{M-1}e^{z_m(\boldsymbol{\omega})}+\tfrac{1}{1-z_M(\boldsymbol{\omega})}\left(\!\prod_{m=1}^{M-1}\tfrac{1}{1-z_m(\boldsymbol{\omega})}\!-\!\prod_{m=1}^{M-1}e^{z_m(\boldsymbol{\omega})}\!\right).
\end{equation}
\else
\begin{multline}
	\label{eq:product_bound_z}
	\!\!\!\!\!\!\prod_{m=1}^M\tfrac{1}{1-z_m(\boldsymbol{\omega})}-\prod_{m=1}^Me^{z_m(\boldsymbol{\omega})}=
	\left(\tfrac{1}{1-z_M(\boldsymbol{\omega})}-e^{z_M(\boldsymbol{\omega})}\right)\times\\
	\prod_{m=1}^{M-1}e^{z_m(\boldsymbol{\omega})}+\tfrac{1}{1-z_M(\boldsymbol{\omega})}\left(\!\prod_{m=1}^{M-1}\tfrac{1}{1-z_m(\boldsymbol{\omega})}\!-\!\prod_{m=1}^{M-1}e^{z_m(\boldsymbol{\omega})}\!\right).
\end{multline}
\fi
Using (\ref{eq:z_less_1}) we obtain:
\if\mycmd1
\begin{equation}
	\label{eq:product_bound_z2}
	\!\!\!\!\!\!\left|\prod_{m=1}^M\frac{1}{1-z_m(\boldsymbol{\omega})}-\prod_{m=1}^Me^{z_m(\boldsymbol{\omega})}\right|\leq
	\left|\frac{1}{1-z_M(\boldsymbol{\omega})}-e^{z_M(\boldsymbol{\omega})}\right|+\left|\prod_{m=1}^{M-1}\frac{1}{1-z_m(\boldsymbol{\omega})}\!-\!\prod_{m=1}^{M-1}e^{z_m(\boldsymbol{\omega})}\right|.
\end{equation}
\else
\begin{multline}
	\label{eq:product_bound_z2}
	\!\!\!\!\!\!\left|\prod_{m=1}^M\frac{1}{1-z_m(\boldsymbol{\omega})}-\prod_{m=1}^Me^{z_m(\boldsymbol{\omega})}\right|\leq
	\left|\frac{1}{1-z_M(\boldsymbol{\omega})}-e^{z_M(\boldsymbol{\omega})}\right|\\
	+\left|\prod_{m=1}^{M-1}\frac{1}{1-z_m(\boldsymbol{\omega})}\!-\!\prod_{m=1}^{M-1}e^{z_m(\boldsymbol{\omega})}\right|.
\end{multline}
\fi
Repeating this argument $M$ times we get:
\begin{equation}
	\label{eq:product_bound_z3}
	\left|\prod_{m=1}^M\frac{1}{1\!-\! z_m(\boldsymbol{\omega})}\!-\!\prod_{m=1}^Me^{z_m(\boldsymbol{\omega})}\right|\!\leq\!\sum_{m=1}^M\left|\frac{1}{1\!-\! z_m(\boldsymbol{\omega})}-e^{z_m(\boldsymbol{\omega})}\right|.
\end{equation}
\begin{equation}
	\left|\frac{1}{1-z_m(\boldsymbol{\omega})}-e^{z_m(\boldsymbol{\omega})}\right|\leq 2.
	\label{eq:first_bound_z}
\end{equation}
Also, using (\ref{eq:z_less_1})  we can write:
\if\mycmd1
\begin{equation}
	\label{eq:second_bound_z}
	\left|\frac{1}{1-z_m(\boldsymbol{\omega})}-e^{z_m(\boldsymbol{\omega})}\right|=\left|\frac{1}{1-z_m(\boldsymbol{\omega})}\right|\left|e^{z_m(\boldsymbol{\omega})}\right|\left|1-z_m(\boldsymbol{\omega})-e^{-z_m(\boldsymbol{\omega})}\right|
	\leq\left|1-z_m(\boldsymbol{\omega})-e^{-z_m(\boldsymbol{\omega})}\right|.
\end{equation}
\else
\begin{eqnarray}
	\label{eq:second_bound_z}
	\left|\frac{1}{1-z_m(\boldsymbol{\omega})}-e^{z_m(\boldsymbol{\omega})}\right|&=&\left|\frac{1}{1-z_m(\boldsymbol{\omega})}\right|\left|e^{z_m(\boldsymbol{\omega})}\right|\nonumber\\
	&&\times\left|1-z_m(\boldsymbol{\omega})-e^{-z_m(\boldsymbol{\omega})}\right|\nonumber\\
	&\leq&\left|1-z_m(\boldsymbol{\omega})-e^{-z_m(\boldsymbol{\omega})}\right|.
\end{eqnarray}
\fi
Then, combining (\ref{eq:product_bound_z3}), (\ref{eq:second_bound_z}) and using that for each $m\in[1:M ]$ and $\boldsymbol{\omega}\in\mathbb{R}$ we have $\left|\frac{1}{1-z_m(\boldsymbol{\omega})}-e^{z_m(\boldsymbol{\omega})}\right|\leq 2$, we get the desired result.

\section{Proof of Lemma \ref{lemma:exp}}
\label{ap:exp}
Consider $z\in\mathbb{C}$ such as $\mathfrak{Re}(z)\leq 0$ and $e^{-tz}$ with $t\in[0:1]$. It is easy to show that:
\begin{equation}
	z\int_{0}^1 e^{-tz}dt=1-e^{-z}.
	\label{eq:exp_aux}
\end{equation}
At the same time we have:
\begin{equation}
	\left|\int_{0}^1 e^{-tz}dt\right|\leq \int_0^1 e^{-t\mathfrak{Re}(z)}dt\leq e^{-\mathfrak{Re}(z)},
	\label{eq:exp_aux2}
\end{equation}
where in the last inequality we used that $\mathfrak{Re}(z)\leq 0$. Combining the last two equations we get $|1-e^{-z}|\leq |z| e^{-\mathfrak{Re}(z)}$.  Similarly, we can write:
\begin{equation}
	z\int_0^1 (e^{-tz}-1) dt=1-z-e^{-z}
	\label{eq:exp_aux3}
\end{equation}
and from $\mathfrak{Re}(z)\leq 0$ and  $|1-e^{-z}|\leq |z|e^{-\mathfrak{Re}(z)}$:
\begin{equation}
	\int_0^{1}|e^{-tz}-1|dt \leq |z|e^{-\mathfrak{Re}(z)}.
	\label{eq:exp_aux4}
\end{equation}
Finally, using (\ref{eq:exp_aux3}) and (\ref{eq:exp_aux4}) we get the desired result.

\section{Proof of Theorem \ref{theo:asymp}}
\label{ap:asymp}
The following bound can be trivially derived:
\if\mycmd1
\begin{multline}
	\int_{\mathbb{R}^N}\Big|\Psi^1(\boldsymbol{\omega}|\mathbf{h})-\hat{\Psi}^1(\boldsymbol{\omega}|\mathbf{h})\Big|d\boldsymbol{\omega}\leq
	\min\left\{2M\int_{\mathbb{R}^N}\prod_{n=1}^N \left|1-j\frac{\sigma_v^2\omega_n}{\beta(M)}\right|^{-M}d\boldsymbol{\omega},\right.\\
	\left. \frac{\delta'(\mathbf{h})M}{\beta^2(M)}\int_{\mathbb{R}^N}\sum_{p=1}^N \!\sum_{r=1}^N \!\frac{\prod_{n=1}^N \left|1-j\frac{\sigma_v^2\omega_n}{\beta(M)}\right|^{-M}|\omega_p||\omega_r|}{\sqrt{1\!+\!\frac{\sigma_v^4\omega_p^2}{\beta^2(M)}}\sqrt{1\!+\!\frac{\sigma_v^4\omega_r^2}{\beta^2(M)}}}d\boldsymbol{\omega}\right\}.
	\label{eq:bound_min_int}
\end{multline}
\else
\begin{multline}
	\label{eq:bound_min_int}
	\int_{\mathbb{R}^N}\Big|\Psi^1(\boldsymbol{\omega}|\mathbf{h})-\hat{\Psi}^1(\boldsymbol{\omega}|\mathbf{h})\Big|d\boldsymbol{\omega}\leq \\
	\min\left\{2M\int_{\mathbb{R}^N}\prod_{n=1}^N \left|1-j\frac{\sigma_v^2\omega_n}{\beta(M)}\right|^{-M}d\boldsymbol{\omega},\right.\\
	\left. \frac{\delta'(\mathbf{h})M}{\beta^2(M)}\int_{\mathbb{R}^N}\sum_{p=1}^N \!\sum_{r=1}^N \!\frac{\prod_{n=1}^N \left|1-j\frac{\sigma_v^2\omega_n}{\beta(M)}\right|^{-M}|\omega_p||\omega_r|}{\sqrt{1\!+\!\frac{\sigma_v^4\omega_p^2}{\beta^2(M)}}\sqrt{1\!+\!\frac{\sigma_v^4\omega_r^2}{\beta^2(M)}}}d\boldsymbol{\omega}\right\}.
\end{multline}
\fi
We will mainly analyze the second term in the $\min$ operator in the RHS of the above equation. The first term can be easily analyzed with the same arguments that will be given in the following. First notice that the mentioned second term can be cast as the sum of $N^2$ multidimensional integrals. Each one of those integrals can be written as:
\begin{equation}
	\int_{\mathbb{R}^N}\left(\prod_{n=1}^{N}\frac{1}{\left(1+\frac{\sigma_v^4\omega_n^2}{\beta^2(M)}\right)^{M/2}}\right)\frac{|\omega_p||\omega_r|}{\sqrt{1\!+\!\frac{\sigma_v^4\omega_p^2}{\beta^2(M)}}\sqrt{1\!+\!\frac{\sigma_v^4\omega_r^2}{\beta^2(M)}}} d\boldsymbol{\omega},
	\label{eq:integrals}
\end{equation}
for each $p,r\in[1:N]$. We need to analyze two cases:
\subsubsection{Case 1: $p\neq r$}
In this case (\ref{eq:integrals}) reduces to:
\begin{equation}
	\left(\!\int_{-\infty}^\infty \!\!\frac{1}{\left(1\!+\!\frac{\sigma_v^4\omega^2}{\beta^2(M)}\right)^{M/2}} d\omega\!\right)^{N-2}\!\!\!\left(\!\int_{-\infty}^\infty\!\! \frac{|\omega|}{\left(1\!+\!\frac{\sigma_v^4\omega^2}{\beta^2(M)}\right)^{\frac{M+1}{2}}} d\omega\!\right)^2
	\label{eq:integrals_p_neq_r}
\end{equation}
\subsubsection{Case 1: $p= r$}
In this case (\ref{eq:integrals}) reduces to:
\begin{equation}
	\left(\int_{-\infty}^\infty \frac{1}{\left(1\!+\!\frac{\sigma_v^4\omega^2}{\beta^2(M)}\right)^{M/2}} d\omega\right)^{N-1}\!\!\!\int_{-\infty}^\infty \frac{\omega^2}{\left(1\!+\!\frac{\sigma_v^4\omega^2}{\beta^2(M)}\right)^{\frac{M+2}{2}}} d\omega.
	\label{eq:integrals_p_eq_r}
\end{equation}
From (\ref{eq:integrals_p_neq_r}) and (\ref{eq:integrals_p_eq_r}), we see that the calculation of (\ref{eq:bound_min_int}) depends on the values of  3 different one-dimensional improper integrals. In first place we consider:
\if\mycmd1
\begin{equation}
	\label{eq:first_integral}
	\int_{-\infty}^\infty\!\! \frac{|\omega|}{\left(1\!+\!\frac{\sigma_v^4\omega^2}{\beta^2(M)}\right)^{\frac{M+1}{2}}} d\omega%= 2\int_{0}^\infty\!\! \frac{|\omega|}{\left(1\!+\!\frac{\sigma_v^4\omega^2}{\beta^2(M)}\right)^{\frac{M+1}{2}}}d\omega
	= \frac{2\beta^2(M)}{\sigma_v^4 (M-1)},
\end{equation}
\else
\begin{eqnarray}
	\label{eq:first_integral}
	\int_{-\infty}^\infty\!\! \frac{|\omega|}{\left(1\!+\!\frac{\sigma_v^4\omega^2}{\beta^2(M)}\right)^{\frac{M+1}{2}}} d\omega%&=& 2\int_{0}^\infty\!\! \frac{|\omega|}{\left(1\!+\!\frac{\sigma_v^4\omega^2}{\beta^2(M)}\right)^{\frac{M+1}{2}}}d\omega\nonumber\\
	&=& \frac{2\beta^2(M)}{\sigma_v^4 (M-1)},
\end{eqnarray}
\fi
where the final value is easily obtained from simple change of variables. In second place from \cite{Gradshteyn_Ryzhik_Jeffrey_Zwillinger_2000} (Equation 3.251) we can obtain\footnote{We assume that $M$ is even. Using complex contour integration we can obtain results when $M$ is odd. However, as we are interesting in the results when $M$ is large and in order to save space, we only consider this case.}:
\if\mycmd1
\begin{eqnarray}
	\label{eq:second_integral}
	\int_{-\infty}^\infty\!\! \frac{1}{\left(1\!+\!\frac{\sigma_v^4\omega^2}{\beta^2(M)}\right)^{\frac{M}{2}}} d\omega%= 2\int_{0}^\infty\!\! \frac{1}{\left(1\!+\!\frac{\sigma_v^4\omega^2}{\beta^2(M)}\right)^{\frac{M}{2}}}
	=\frac{(M-3)!!\beta(M)\pi}{(M-2)!!\sigma_v^2},
\end{eqnarray}
\else
\begin{eqnarray}
	\label{eq:second_integral}
	\int_{-\infty}^\infty\!\! \frac{1}{\left(1\!+\!\frac{\sigma_v^4\omega^2}{\beta^2(M)}\right)^{\frac{M}{2}}} d\omega%&=& 2\int_{0}^\infty\!\! \frac{1}{\left(1\!+\!\frac{\sigma_v^4\omega^2}{\beta^2(M)}\right)^{\frac{M}{2}}}\nonumber\\
	&=& \frac{(M-3)!!\beta(M)\pi}{(M-2)!!\sigma_v^2},
\end{eqnarray}
\fi
and
\if\mycmd1
\begin{eqnarray}
	\label{eq:third_integral}
	\int_{-\infty}^\infty\!\! \frac{\omega^2}{\left(1\!+\!\frac{\sigma_v^4\omega^2}{\beta^2(M)}\right)^{\frac{M+2}{2}}} d\omega%= 2\int_{0}^\infty\!\! \frac{\omega^2}{\left(1\!+\!\frac{\sigma_v^4\omega^2}{\beta^2(M)}\right)^{\frac{M+2}{2}}}
	= \frac{(M-3)!!\beta^3(M)\pi}{(M)!!\sigma_v^6},
\end{eqnarray}
\else
\begin{eqnarray}
	\label{eq:third_integral}
	\int_{-\infty}^\infty\!\! \frac{\omega^2}{\left(1\!+\!\frac{\sigma_v^4\omega^2}{\beta^2(M)}\right)^{\frac{M+2}{2}}} d\omega%&=& 2\int_{0}^\infty\!\! \frac{\omega^2}{\left(1\!+\!\frac{\sigma_v^4\omega^2}{\beta^2(M)}\right)^{\frac{M+2}{2}}}\nonumber\\
	&=& \frac{(M-3)!!\beta^3(M)\pi}{(M)!!\sigma_v^6},
\end{eqnarray}
\fi
where $n!!$ for $n\in\mathbb{N}$ is the double factorial of $n$. It is well known that when $n=2l$ with $l\in\mathbb{N}$, then $n!!=2^l l!$ and when $n=2l-1$, $n!!=\frac{(2l)!}{2^l l!}$. Using those facts, combining all the above results, and using Stirling approximation we can obtain the following bound for  the second term in the RHS in (\ref{eq:bound_min_int}):
\if\mycmd1
\begin{equation}
	\int_{\mathbb{R}^N}\sum_{p=1}^N \!\sum_{r=1}^N \!\frac{\prod_{n=1}^N \left|1-j\frac{\sigma_v^2\omega_n}{\beta(M)}\right|^{-M}|\omega_p||\omega_r|}{\sqrt{1\!+\!\frac{\sigma_v^4\omega_p^2}{\beta^2(M)}}\sqrt{1\!+\!\frac{\sigma_v^4\omega_r^2}{\beta^2(M)}}}d\boldsymbol{\omega}=
	\mathcal{O}\left(\frac{\beta^{N+2}(M)}{M^{\frac{N+2}{2}}}\right),
\end{equation}
\else
\begin{multline}
	\int_{\mathbb{R}^N}\sum_{p=1}^N \!\sum_{r=1}^N \!\frac{\prod_{n=1}^N \left|1-j\frac{\sigma_v^2\omega_n}{\beta(M)}\right|^{-M}|\omega_p||\omega_r|}{\sqrt{1\!+\!\frac{\sigma_v^4\omega_p^2}{\beta^2(M)}}\sqrt{1\!+\!\frac{\sigma_v^4\omega_r^2}{\beta^2(M)}}}d\boldsymbol{\omega}=\\
	\mathcal{O}\left(\frac{\beta^{N+2}(M)}{M^{\frac{N+2}{2}}}\right),
\end{multline}
\fi
%where $K(N,\sigma_v^2)$ is constant that depends only on $\sigma_v^2$ and $N$.
Finally we can get:
\if\mycmd1
\begin{equation}
	\label{eq:finalbound}
	\frac{1}{(2\pi)^N}\int_{\mathbb{R}^N}\Big|\Psi^1(\boldsymbol{\omega}|\mathbf{h})-\hat{\Psi}^1(\boldsymbol{\omega}|\mathbf{h})\Big|d\boldsymbol{\omega}=
	\mathcal{O}\left(
	\min\left\{\frac{\beta^{N}(M)}{M^{\frac{N}{2}-1}},\delta'(\mathbf{h})\frac{\beta^{N}(M)}{M^{\frac{N}{2}}}\right\}\right).
\end{equation}
\else
\begin{multline}
	\label{eq:finalbound}
	\frac{1}{(2\pi)^N}\int_{\mathbb{R}^N}\Big|\Psi^1(\boldsymbol{\omega}|\mathbf{h})-\hat{\Psi}^1(\boldsymbol{\omega}|\mathbf{h})\Big|d\boldsymbol{\omega}=\\
	\mathcal{O}\left(
	\min\left\{\frac{\beta^{N}(M)}{M^{\frac{N}{2}-1}},\delta'(\mathbf{h})\frac{\beta^{N}(M)}{M^{\frac{N}{2}}}\right\}\right).
\end{multline}
\fi
from which and (\ref{eq:CF_bound_max}) the desired result follows.
%\bibliographystyle{IEEEtran}
%\bibliography{./refs}
% Bibliography (copy paste after accepted)

\end{document}